\pdfoutput=1
\newcommand*{\ATLASLATEXPATH}{latex/}
\documentclass[UKenglish,cernpreprint,texlive=2016]{\ATLASLATEXPATH atlasdoc}
\usepackage[subfigure]{\ATLASLATEXPATH atlaspackage}
\usepackage{\ATLASLATEXPATH atlasbiblatex}
\usepackage{dcolumn}
\usepackage{multirow}
\usepackage{enumerate}
\usepackage [BSM, other,process,particle]{\ATLASLATEXPATH atlasphysics}
\graphicspath{{logos/}}
\AtlasTitle{Searches for exclusive Higgs and $Z$ boson decays into $\Jpsi\,\gamma$, $\psi(2S)\,\gamma$, and $\Upsilon(nS)\,\gamma$ at $\sqrt{s}=13~\TeV$ with the ATLAS detector}

\author{The ATLAS Collaboration}

\date{\today}

\PreprintIdNumber{CERN-EP-2018-154}

\AtlasJournalRef{Phys. Lett. B 786 (2018) 134}

\AtlasDOI{10.1016/j.physletb.2018.09.024}

\AtlasAbstract{
Searches for the exclusive decays of the Higgs and $Z$ bosons into
a $\Jpsi$, $\psi(2S)$, or $\Upsilon(nS)$ ($n=1,2,3$) meson and a photon are performed with
a $pp$ collision data sample corresponding to an integrated luminosity of
$36.1~\ifb$ collected at $\sqrt{s}=13~\TeV$ with the ATLAS detector at
the CERN Large Hadron Collider.
No significant excess of events is observed above the expected backgrounds, and
95\% confidence-level upper limits on the branching fractions of the
Higgs boson decays to $\Jpsi\,\gamma$, $\psi(2S)\,\gamma$, and $\Upsilon(nS)\,\gamma$ of
$3.5\times10^{-4}$, $2.0\times10^{-3}$, and $(4.9,5.9,5.7)\times10^{-4}$, respectively, are
obtained assuming Standard Model production. The corresponding 95\% confidence-level upper limits for the
branching fractions of the $Z$ boson decays are $2.3\times10^{-6}$, $4.5\times10^{-6}$ and $(2.8,1.7,4.8)\times10^{-6}$, respectively.
}

\hypersetup{pdftitle={ATLAS document},pdfauthor={The ATLAS Collaboration}}

\newcommand\progname[1]{{\scshape #1}}
\def\ee{\ensuremath{e^+ e^-}}

\def\epem{\ensuremath{e^+ e^-}}
\def\mumu{\ensuremath{\mathrm{\mu^+ \mu^-}}}
\def\mm{\ensuremath{\mathrm{\mu\mu}}}
\def\tautau{\ensuremath{\mathrm{\tau^+ \tau^-}}}

\def\leplep{\ensuremath{\ell^+ \ell^-}}
\def\lnu{\ensuremath{\ell \nu}}
\def\ptmumu {\ensuremath{p_{\textrm{T}}^{\,\mm}}}

\def\ptmu {\ensuremath{p_{\textrm{T}}^{\,\mu}}}
\def\ptmuq#1 {\ensuremath{p_{\textrm{T}}^{\,\mu^{#1}}}}
\def\ptmumu {\ensuremath{p_{\textrm{T}}^{\mm}}}
\def\etamu {\ensuremath{\eta^{\,\mu}}}
\def\etamuq#1 {\ensuremath{\eta^{\,\mu^{#1}}}}

\def\ptgamma {\ensuremath{p_{\textrm{T}}^{\gamma}}}
\def\etagamma {\ensuremath{\eta^{\gamma}}}

\def\HtoJpsiGamma{\ensuremath{H\to\Jpsi\,\gamma}}
\def\HtoPsiGamma{\ensuremath{H\to\psi(2S)\,\gamma}}

\newcolumntype{d}[1]{D{.}{\pm}{#1}}
\newcolumntype{h}[1]{D{.}{}{#1}}

\newcommand{\tabscript}[3]{
  \setlength{\fboxrule}{0pt}
  \fbox{\ensuremath{#1^{#2}_{#3}}}

  }
\begin{document}

\maketitle

\section{Introduction}
\label{sec:intro}

Following the observation of a Higgs boson $H$ with a mass of approximately
125~\GeV\ by the ATLAS and CMS
collaborations~\cite{ATLAS:2012af,Chatrchyan:2012ia}, detailed
measurements of its properties show no deviations from the Standard Model (SM)~\cite{CMS-HIG-15-002}. However, its
role in fermion mass generation is still to be shown experimentally. In the SM , this mass generation is implemented through Yukawa interactions,
and many theories beyond the SM predict
substantial modifications of the relevant Higgs boson couplings to
fermions.
The ATLAS and CMS collaborations have reported measurements of the Higgs coupling to a third-generation fermion with a significance greater than five standard deviations in the $H\to\tau^+\tau^-$
channel~\cite{CMS-HIG-15-002,HIGG-2013-32,CMS-HIG-13-004,}. In addition, progress has been made in the third-generation quark sector with indirect evidence of the coupling of the Higgs boson to the top quark~\cite{CMS-HIG-15-002}. This was recently complemented by direct observation of the associated production of the Higgs boson with a top-quark pair ($t\tbar H$)~\cite{Sirunyan:2018hoz,Aaboud:2018urx}. Evidence of Higgs boson decays into $b\bar{b}$ has also been found with a significance in excess of three standard deviations by both ATLAS and CMS~\cite{Aaboud:2017xsd,CMS-HIG-16-044,Aaltonen:2012qt}. No experimental evidence of Higgs boson decays into the first- and second-generation fermions has yet been found, but direct searches were recently performed by the ATLAS Collaboration for $H\to\ccbar$~\cite{Aaboud:2018fhh} and $H\to\mumu$~\cite{HIGG-2016-10,HIGG-2013-07} and by the CMS Collaboration for $H\to\mumu$ and $H\to\ee$ decays~\cite{CMS-HIG-13-007}.

The Standard Model Higgs boson decays $\HtoJpsiGamma$ and $\HtoPsiGamma$ offer an
opportunity to access the $c$-quark Yukawa coupling~\cite{HBranching,VatoHBranching}
in a manner complementary to studies of the inclusive decay $H\to\ccbar$.
The branching fraction for $\HtoJpsiGamma$ has been calculated within the SM to be
${\cal B}(H\to J/\psi\,\gamma) = (2.99^{+0.16}_{-0.15})\times
10^{-6}$~\cite{Bodwin:2017wdu}. Other recent results on these calculations are given in Refs.~\cite{Bodwin:2016edd,Koenig:2015pha,Bodwin:2014bpa}.
For $H\to\psi(2S)\,\gamma$ the branching fraction was calculated by the authors of
Ref.~\cite{Bodwin:2017wdu} to be ${\cal B}(H\to\psi(2S)\,\gamma)=(1.03\pm 0.06)\times 10^{-6}$ using
an estimate for the value of the order-$v^2$ NRQCD long-distance
matrix element.

The corresponding decays in the bottomonium sector, $H\to\Upsilon(1S,2S,3S)\,\gamma$,
can provide, in combination with $H\to b\bar{b}$ decays, information about the real and imaginary parts
of the $b$-quark coupling to the Higgs boson~\cite{Koenig:2015pha}, which could probe potential CP violation in the Higgs sector. However, the expected
SM branching fractions,
${\cal B}(H\to \Upsilon(nS)\,\gamma) = (5.22^{+2.02}_{-1.70},1.42^{+0.72}_{-0.57},0.91^{+0.48}_{-0.38})\times10^{-9}$ ($n=1,2,3$)~\cite{Bodwin:2016edd,Bodwin:2017wdu}, are smaller due to a cancellation between the ``direct'' and ``indirect'' amplitudes. The direct amplitude proceeds through the $H\to q\bar{q}$ coupling with a subsequent photon emission
before the $q\bar{q}$ hadronisation to $\Upsilon(nS)$. The indirect amplitude proceeds via the
$H\gamma\gamma$ coupling followed by the fragmentation $\gamma^*\to \Upsilon(nS)$.

Deviations of the $c$- and $b$-quark Yukawa couplings from the SM expectations can
lead to significant increases in the branching fractions for exclusive decays. These deviations can arise in beyond-the-SM theories; for example, the quark masses might not originate entirely from the Higgs mechanism but could also be induced by other subdominant sources of electroweak symmetry breaking~\cite{Perez:2015aoa}. Other scenarios include the minimal flavour violation
framework~\cite{DAmbrosio:2002ex}, the Froggatt--Nielsen
mechanism~\cite{Froggatt:1978nt}, the Higgs-dependent Yukawa couplings
model~\cite{Giudice:2008uua}, the Randall--Sundrum family of models
\cite{Randall:1999ee}, and the possibility of the Higgs boson being a
composite pseudo-Goldstone boson~\cite{Dugan:1984hq}. An overview of
relevant models of physics beyond-the-SM is provided in
Ref.~\cite{deFlorian:2016spz}.

Decays of the Higgs boson into a heavy, vector quarkonium
state, ${\cal Q}\equiv\Jpsi$ or $\Upsilon(nS)$, and a photon
were searched for by the ATLAS
Collaboration with
up to $19.2~\ifb$ of data collected at $\sqrt{s}=8~\TeV$~\cite{HIGG-2014-03}, resulting in 95\% confidence level (CL)
upper limits of $1.5\times 10^{-3}$ for ${\cal B}\left(\HtoJpsiGamma\right)$ and
$\left(1.3,1.9,1.3\right)\times 10^{-3}$ for ${\cal B}\left(H\to\Upsilon(nS)\,\gamma\right)$ ($n=1,2,3$).
The former decay mode was also searched for by the CMS
Collaboration~\cite{CMS-HIG-14-003}, yielding a similar upper limit.
In addition, the ATLAS Collaboration searched for the rare Higgs decays
$H\to\phi\,\gamma$ and $H\to\rho\,\gamma$~\cite{HIGG-2016-05,HIGG-2016-13}.

Owing to the large $Z$ boson production cross section at the LHC, rare
$Z$ boson decays can be probed at much lower rates than for Higgs
boson decays into the same final state. Branching fractions for
$Z\to{\cal Q}\,\gamma$ decays have been calculated to be between
$10^{-8}$ and $10^{-7}$ for both the $Z\to\Jpsi\,\gamma$ and
$Z\to\Upsilon(nS)\,\gamma$
decays~\cite{Bodwin:2017pzj,Grossmann:2015lea,Huang:2014cxa}. Measurements of the branching
fractions for these decays would provide a sensitive test of the SM and the
factorisation approach in quantum
chromodynamics (QCD), since the power
corrections in terms of the ratio of the QCD energy scale to the vector-boson
mass are small~\cite{Grossmann:2015lea}.
ATLAS searched for $Z$ boson decays into
$\Jpsi$ or $\Upsilon(nS)$ ($n=1,2,3$) and a photon with $20.3~\ifb$ of data collected at $\sqrt{s}=8~\TeV$~\cite{HIGG-2014-03},
resulting in 95\% CL upper limits of $2.6\times 10^{-6}$ and
$\left(3.4,6.5,5.4\right)\times 10^{-6}$, respectively.
ATLAS also searched for the decay modes $Z\to\phi\,\gamma$ and $Z\to\rho\,\gamma$~\cite{HIGG-2016-05,HIGG-2016-13}.

This Letter describes searches for Higgs and $Z$ boson decays into the
exclusive final states $\Jpsi\,\gamma$, $\psi(2S)\,\gamma$, and $\Upsilon(nS)\,\gamma$ ($n=1,2,3$) with $\Jpsi\to\mumu$, $\psi(2S)\to\mumu$, and $\Upsilon(nS)\to\mumu$
using ATLAS data collected in 2015 and 2016 at $\rts=13~\TeV$.
Throughout the remainder of this Letter, where no distinction is relevant, the $\Jpsi$ and $\psi(2S)$ states are referred to collectively as the $\psi(nS)$ states.

\section{ATLAS detector}
\label{sec:atlas}

ATLAS~\cite{Aad:2008zzm} is a multipurpose particle
detector with a forward--backward symmetric cylindrical
geometry and near 4$\pi$ coverage in solid angle.\footnote{ATLAS uses a
  right-handed coordinate system with its origin at the nominal
  interaction point (IP) in the centre of the detector and the
  $z$-axis along the beam pipe. The $x$-axis points from the IP to the
  centre of the LHC ring, and the $y$-axis points upward. Cylindrical
  coordinates $(r,\phi)$ are used in the transverse plane, $\phi$
  being the azimuthal angle around the $z$-axis. The pseudorapidity is
  defined in terms of the polar angle $\theta$ as
  $\eta=-\ln\tan(\theta/2)$.} It consists of an inner tracking detector, electromagnetic and hadronic calorimeters, and a muon spectrometer.

The inner tracking detector (ID) covers the pseudorapidity range
$|\eta|<2.5$ and is surrounded by a thin superconducting solenoid
providing a 2~T magnetic field.
At small radii, a high-granularity silicon pixel detector surrounds the
vertex region and typically provides four measurements per track. It is followed by a silicon microstrip tracker, which provides eight
measurement points per track. The silicon detectors are complemented
by a gas-filled straw-tube transition radiation tracker, which enables
radially extended track reconstruction up to $|\eta| = 2.0$ with
typically 35 measurements per track.

Electromagnetic (EM) calorimetry within the region $|\eta|<3.2$ is
provided by barrel and endcap high-granularity lead/liquid-argon (LAr)
EM calorimeters with an additional thin LAr presampler covering
$|\eta| < 1.8$ to correct for energy loss in upstream material; for $|\eta| < 2.5$ the EM calorimeter is divided into
three layers in depth.
A steel/scintillator-tile calorimeter provides hadronic calorimetry for $|\eta|<1.7$. LAr technology with copper as absorber is
used for the hadronic calorimeters in the endcap region
$1.5<|\eta|<3.2$.  The solid-angle coverage is completed with forward
copper/LAr and tungsten/LAr calorimeter modules in $3.1 < |\eta| <
4.9$ optimised for EM and hadronic measurements,
respectively.

The muon spectrometer surrounds the calorimeters and has
separate trigger and high-precision tracking chambers measuring the
deflection of muons in a magnetic field provided by three air-core
superconducting toroidal magnets. The precision chamber system covers the
region $|\eta| < 2.7$ with three layers of monitored drift tubes,
complemented by cathode strip chambers in the forward region. The muon
trigger system covers the range $|\eta| < 2.4$ with resistive plate
chambers in the barrel and thin gap chambers in the endcap regions.

A two-level trigger and data acquisition system is used to record
events for offline analysis~\cite{TRIG-2016-01}. The level-1 trigger
is implemented in hardware and uses a subset of detector information
to reduce the event rate to at most 100~kHz. It is followed
by a software-based high-level trigger which filters events using the
full detector information and outputs events for permanent storage at
an average rate of 1~kHz.

\section{Data and simulated data}
\label{sec:datamc}

The search is performed with a sample of $pp$ collision data recorded
at a centre-of-mass energy $\rts=13~\TeV$ corresponding to an
integrated luminosity of 36.1~\ifb. The integrated luminosity of
the data sample has an uncertainty of $2.1\%$ derived using the method
described in Ref.~\cite{DAPR-2013-01}. Events are retained for
further analysis only if they were collected under stable LHC beam
conditions and all relevant detector components were fully functional.

The data samples used in this analysis were collected by a combination
of two triggers~\cite{TRIG-2016-01}. The first required an isolated photon with transverse
momentum $\ptgamma$ greater than $35~\GeV$ and at least one muon with
$\ptmu$ greater than $18~\GeV$. The second required an isolated photon
with $\ptgamma>25~\GeV$ and a muon with $\ptmu>24~\GeV$. The use of two triggers with differing transverse momentum thresholds on the photon and muon objects offers an increased trigger efficiency with respect to the case of either trigger used alone.

Higgs~boson production was modelled using the \progname{Powheg-Box}~v2
Monte Carlo (MC) event
generator~\cite{powheg1,powheg2,powheg3,powheg4,powheg5} with CT10 parton distribution
functions~\cite{Lai:2010vv} for the gluon--gluon fusion ($ggH$) and vector-boson fusion (VBF) processes. Both processes were calculated up
to next-to-leading order (NLO) in $\alphas$. \progname{Powheg-Box} was interfaced with
\progname{Pythia\,8.186}~\cite{Sjostrand:2007gs,Sjostrand:2006za} to model the parton shower,
hadronisation, and underlying event with
the AZNLO set of tuned parameters~\cite{STDM-2012-23}. Additional contributions from the associated production of a Higgs
boson and a $W$ or $Z$ boson (denoted $WH$ and $ZH$, respectively) were
modelled by the \progname{Pythia\,8.186} event
generator~\cite{Sjostrand:2007gs,Sjostrand:2006za} with NNPDF23LO
parton distribution functions~\cite{Ball:2012cx} and the A14 set of tuned parameters for hadronisation and the underlying event~\cite{ATL-PHYS-PUB-2014-021}.
The production rates for the SM Higgs boson with $\mH = 125~\GeV$, obtained from the compilation in Ref.~\cite{deFlorian:2016spz}, are assumed
throughout this analysis.
The $ggH$ production is normalised such that it reproduces the total cross section predicted by a next-to-next-to-next-to-leading-order QCD calculation with NLO electroweak corrections applied~\cite{Anastasiou:2015ema,Anastasiou:2016cez,Actis:2008ug,Anastasiou:2008tj}. The VBF production is normalised to an approximate next-to-next-to-leading-order (NNLO) QCD cross section with NLO electroweak corrections applied~\cite{Ciccolini:2007jr,Ciccolini:2007ec,Bolzoni:2010xr}. The production of $WH$ and $ZH$ is normalised to
cross sections calculated at NNLO in QCD with NLO electroweak corrections~\cite{Brein:2003wg,Denner:2011id} including the NLO QCD corrections~\cite{Altenkamp:2012sx} for $gg\to ZH$. The production of a Higgs boson in association with $t\bar{t}$ ($t\bar{t}H$) or $b\bar{b}$ ($b\bar{b}H$) is accounted for by scaling the total cross section used to normalise the $ggH$ signal sample, assuming the signal efficiency of these processes to be equal to that for $ggH$. The addition of $t\bar{t}H$ and $b\bar{b}H$ to the $ggH$ signal changes the acceptance by less than 1\%.

The \progname{Powheg-Box}~v2 event generator was also used to model
inclusive $Z$ boson production.
\progname{Pythia\,8.186}~\cite{Sjostrand:2007gs,Sjostrand:2006za} with
CTEQ6L1 parton distribution functions~\cite{Pumplin:2002vw} and the
AZNLO set of hadronisation and underlying-event
parameters~\cite{STDM-2012-23} was used to simulate parton showering and
hadronisation.
The prediction is normalised to the total cross section obtained from
the measurement in Ref.~\cite{STDM-2015-03}, which has an uncertainty
of 2.9\%. This conservatively assumes the luminosity component of the
uncertainty to be completely uncorrelated with that of this search.

The Higgs and $Z$ boson decays were simulated as a cascade of two-body
decays, accounting for effects of the quarkonium helicity on the
$\mumu$ kinematics. The quarkonium state is simulated to be transversely polarised in the case of the Higgs boson decay and longitudinally polarised in the case of the $Z$ boson decay~\cite{Huang:2014cxa}. The branching fractions for the decays $\cal{Q}\to\mumu$ are taken from Ref.~\cite{PDG}.
The simulated events were passed through the detailed
\GEANT~4 simulation of the ATLAS detector~\cite{GEANT4,SOFT-2010-01} and
processed with the same software used to reconstruct the data.

\section{Event selection for $\psi(nS)\,\gamma\to \mumu\gamma$ and $\Upsilon(nS)\,\gamma\to\mumu\gamma$}
\label{sec:selection}

Muons are reconstructed from ID tracks combined with independent
muon spectrometer tracks or track segments~\cite{PERF-2014-05} and are
required to have $\ptmu>3~\GeV$ and pseudorapidity
$|\etamu|<2.5$. Candidate $\cal{Q}\to\mumu$ decays are reconstructed from
pairs of oppositely charged muons consistent with originating from a
common vertex. The highest-$\pt$ muon in a pair, called the leading muon
in the following, is required to have $\ptmu>18~\GeV$. Dimuons with a
mass $m_\mumu$ within $2.0<m_\mumu<4.2~\GeV$ are identified as $\psi(nS)\to\mumu$ candidates. Dimuons with $8.0<m_\mumu<12.0~\GeV$ are considered to be
$\Upsilon(nS)\to\mumu$ candidates.

Selected $\cal{Q}\to\mumu$ candidates are subjected to isolation and
vertex-quality requirements.
In this case, the primary $pp$ vertex is defined as the reconstructed
vertex with the highest $\Sigma_i{\,\pt}^2_i$ of all associated tracks used to
form the vertex.
The sum of the $\pt$ of the tracks within
$\Delta R = \sqrt{(\Delta\phi)^{2} + (\Delta\eta)^{2}} = 10~\GeV/\ptmu$ (with a maximum $\Delta R$ of 0.3) of the
leading muon is required to be less than 6\% of the muon $\pt$~\cite{PERF-2015-10}. To mitigate the effects of multiple $pp$ interactions in the same or
neighbouring bunch crossings, only ID tracks that originate from the primary vertex are considered.
The transverse momentum of the ID track associated with the
leading muon is subtracted from the sum, and the subleading muon is
also subtracted if it falls within the isolation cone. To reject backgrounds from $b$-hadron decays, the signed projection of the $\cal{Q}$ candidate's flight distance between the primary $pp$ vertex and the dimuon vertex onto the direction of its transverse momentum is required to be less than three times its uncertainty.

Photons are reconstructed from clusters of energy in the
electromagnetic calorimeter. Clusters without matching ID tracks are
classified as unconverted photon candidates while clusters matched to
ID tracks consistent with the hypothesis of a photon conversion into $\ee$
are classified as converted photon candidates~\cite{Aaboud:2016yuq}.  Reconstructed photon
candidates are required to have transverse momentum $\ptgamma>35~\GeV$ and
pseudorapidity $|\etagamma|<2.37$, excluding the barrel/endcap
calorimeter transition region $1.37<|\etagamma|<1.52$, and to satisfy
the ``tight'' photon identification criteria~\cite{Aaboud:2016yuq}.
Isolation requirements are imposed to further suppress the contamination
from jets. The sum of the transverse momenta of all tracks originating from the primary vertex, within
$\Delta R = 0.2$ of the photon direction, excluding those associated
with reconstructed photon conversions, is required to be less than $5\%$ of
$\ptgamma$.
In addition to this track isolation criterion, a calorimeter isolation criterion is applied where the sum of the transverse momenta of calorimeter energy clusters within $\Delta R = 0.4$ of the photon direction, excluding the transverse energy of the reconstructed photon, is required
to be less than $\left(2.45~\GeV + 0.022 \ptgamma\right)$.
The effects
of multiple $pp$ interactions in the events are also accounted for in the calorimeter isolation measurements.

Combinations of a $\cal{Q}\to\mumu$ candidate and a photon satisfying
$\Delta\phi({\cal{Q}},\gamma)>\pi/2$ are retained for further
analysis. When multiple combinations are possible, a situation that
arises in only a few percent of the events, the combination of the
highest-\pt photon and the $\cal {Q}$ candidate with an invariant mass
closest to the respective quarkonium mass is retained. To improve the sensitivity of the $\Upsilon(nS)\gamma$ analysis in resolving the individual $\Upsilon(nS)$ states, the events are classified into two exclusive categories based upon the pseudorapidity of the muons. Events where both muons are within the region $|\etamu|<1.05$ constitute the ``barrel'' (B) category. Events where at least one of the muons is outside the region $|\etamu|<1.05$ constitute the ``endcap'' (EC) category.

To maintain a single search region, while ensuring near-optimal sensitivity for both the Higgs and $Z$ boson analyses, the transverse
momentum of the $\cal {Q}$ candidate $\pt^{\cal Q}$
is required to be greater than a value that varies as a function of
the invariant mass of the three-body system $m_{\cal{Q}\gamma}$. For the $\psi(nS)\to\mumu$ ($\Upsilon(nS)\to\mumu$) selection, $\pt^{\cal{Q}}$ thresholds of
$40~\GeV$ ($34~\GeV$) and $54.4~\GeV$ ($52.7~\GeV$) are imposed for the regions
$m_{\cal{Q}\gamma}\leq91~\GeV$ and $m_{\cal{Q}\gamma}\geq140~\GeV$,
respectively. The thresholds are varied between their minimum and maximum
values as a linear function of $m_{\cal{Q}\gamma}$ in the region
$91~\GeV<m_{\cal{Q}\gamma}<140~\GeV$.

\section{Signal modelling}
\label{sec:signalmodel}
For the $\psi(nS)\,\gamma\to\mumu\,\gamma$ final state, the total signal efficiencies
(kinematic acceptance, trigger, reconstruction, identification, and isolation efficiencies) are
19\% and 11\% for the Higgs and $Z$~boson decays, respectively. The
corresponding efficiencies for the $\Upsilon(nS)\,\gamma\to\mumu\,\gamma$ final states
are 22--23\% and 15--16\%, respectively.
The difference in efficiency between the Higgs and $Z$ boson decays arises
primarily from the softer
$\ptgamma$ and $\ptmumu$ distributions associated with
$Z\to\cal{Q}\,\gamma$ production, as seen by comparing Figures~\ref{fig:pt_ph_muons_Hjpsi} and \ref{fig:pt_ph_muons_Zjpsi} for the $\Jpsi\,\gamma$ case and Figures~\ref{fig:pt_ph_muons_Hupsi} and \ref{fig:pt_ph_muons_Zupsi} for the $\Upsilon(nS)\,\gamma$ case.

Accounting for quarkonium helicity effects in the $\cal{Q}\to\mumu$
decays leads to a 3--4\% decrease of the expected Higgs boson
efficiency and a corresponding 6--9\% increase of the expected $Z$
boson efficiency, relative to the efficiency for an isotropic decay.

\begin{figure}[!h]
\centering
\subfigure[\label{fig:pt_ph_muons_Hjpsi}]{\includegraphics[width=0.40\textwidth]{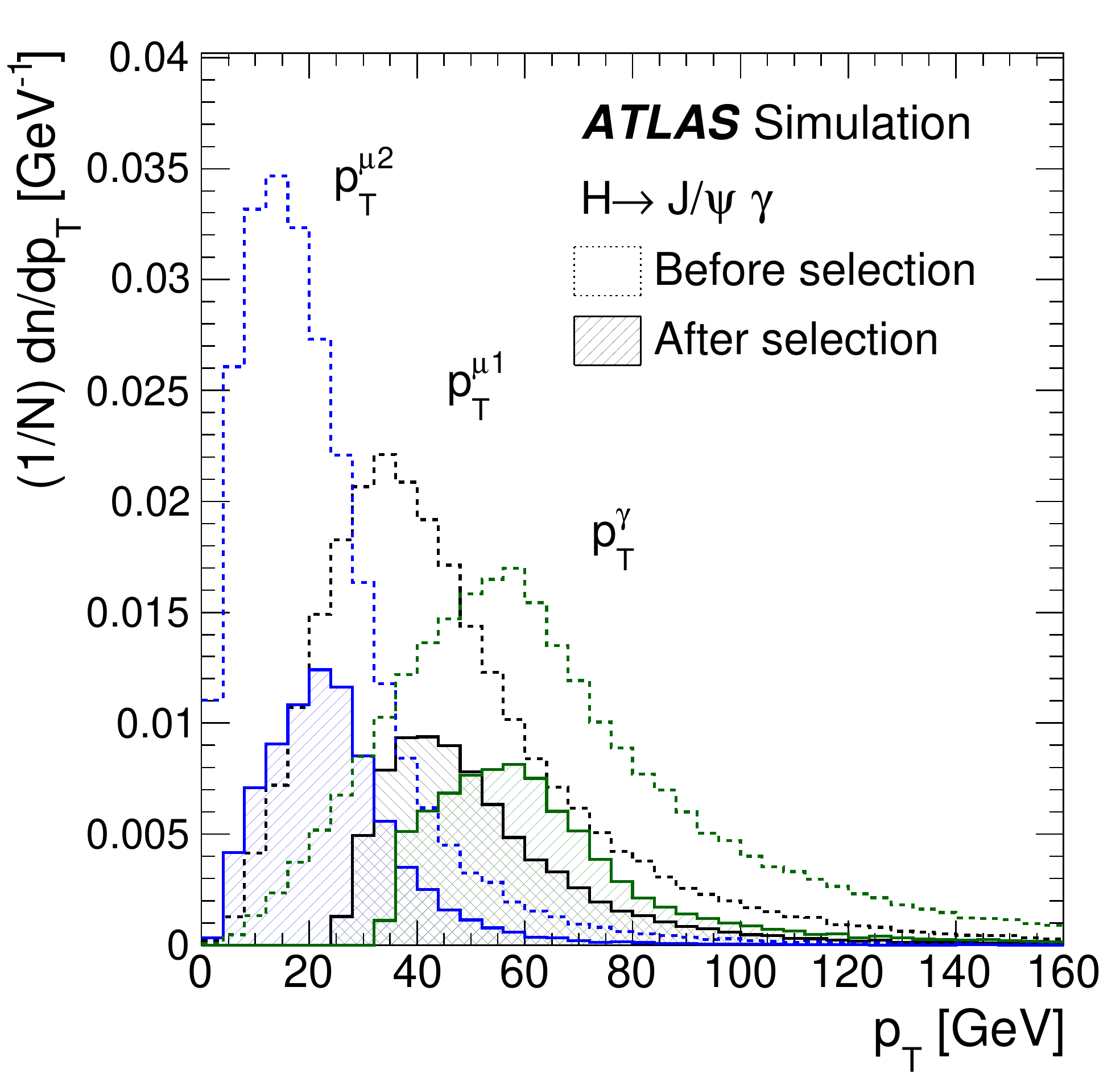}}
\subfigure[\label{fig:pt_ph_muons_Zjpsi}]{\includegraphics[width=0.40\textwidth]{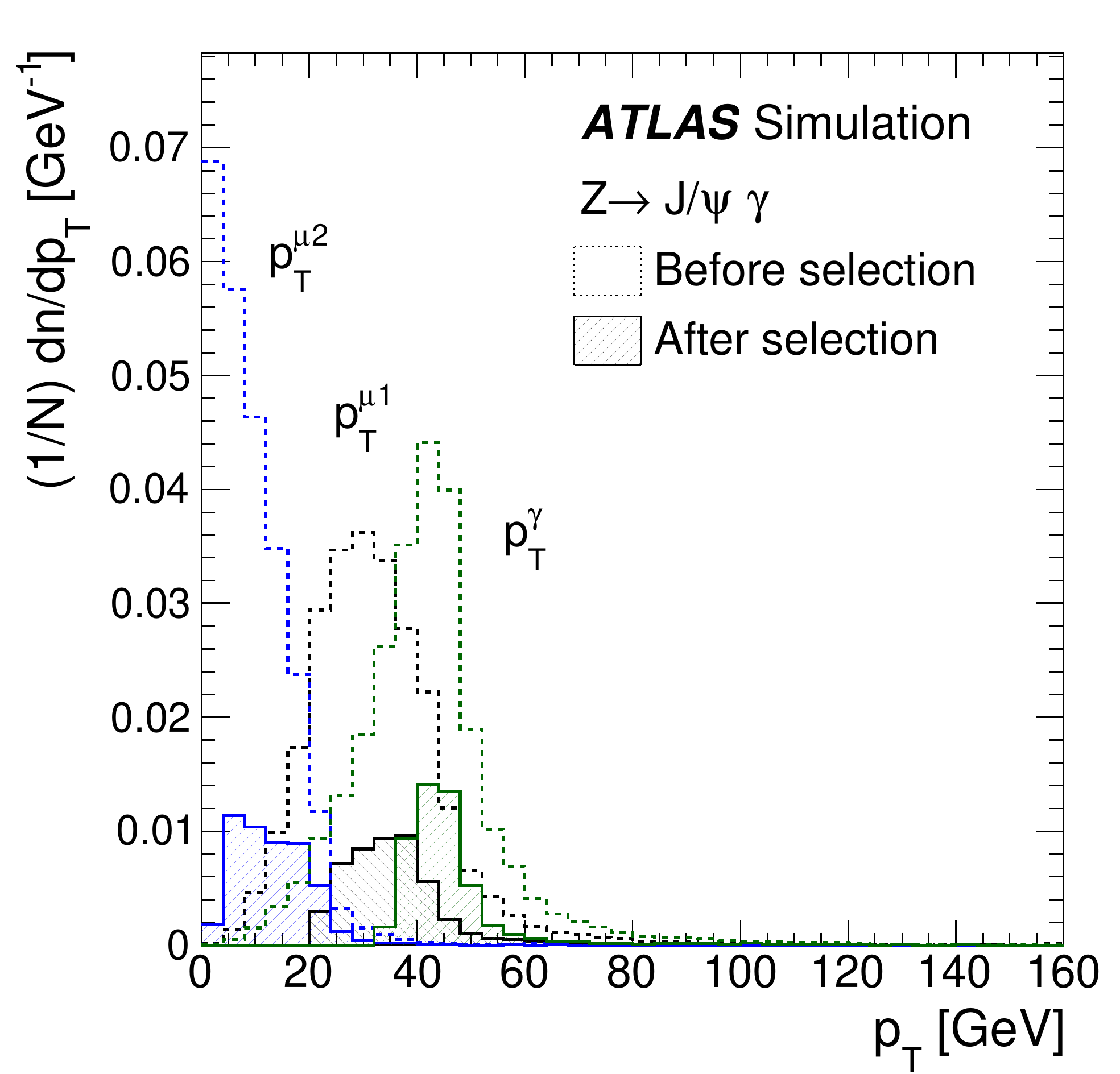}}

\subfigure[\label{fig:pt_ph_muons_Hupsi}]{\includegraphics[width=0.40\textwidth]{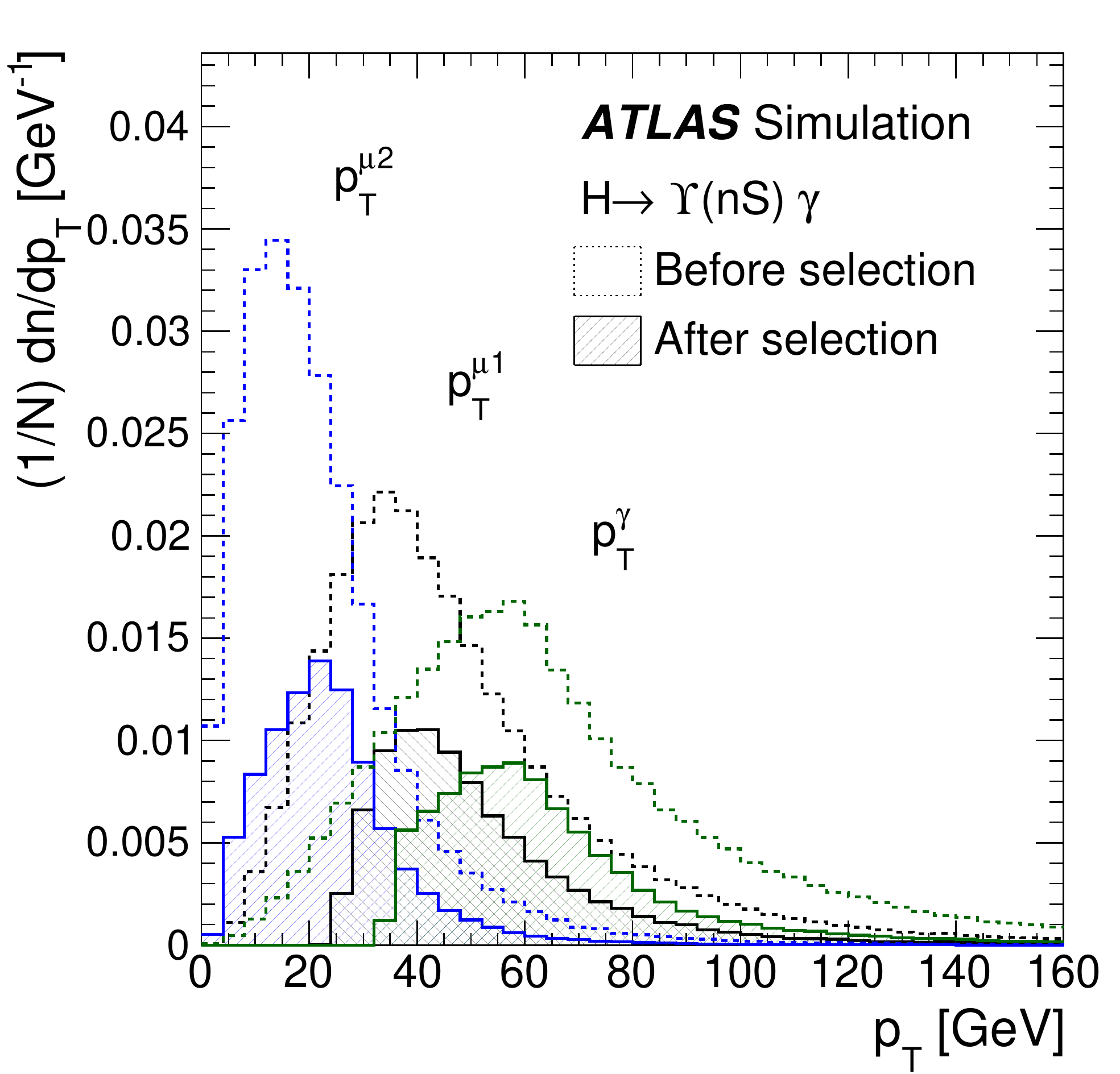}}
\subfigure[\label{fig:pt_ph_muons_Zupsi}]{\includegraphics[width=0.40\textwidth]{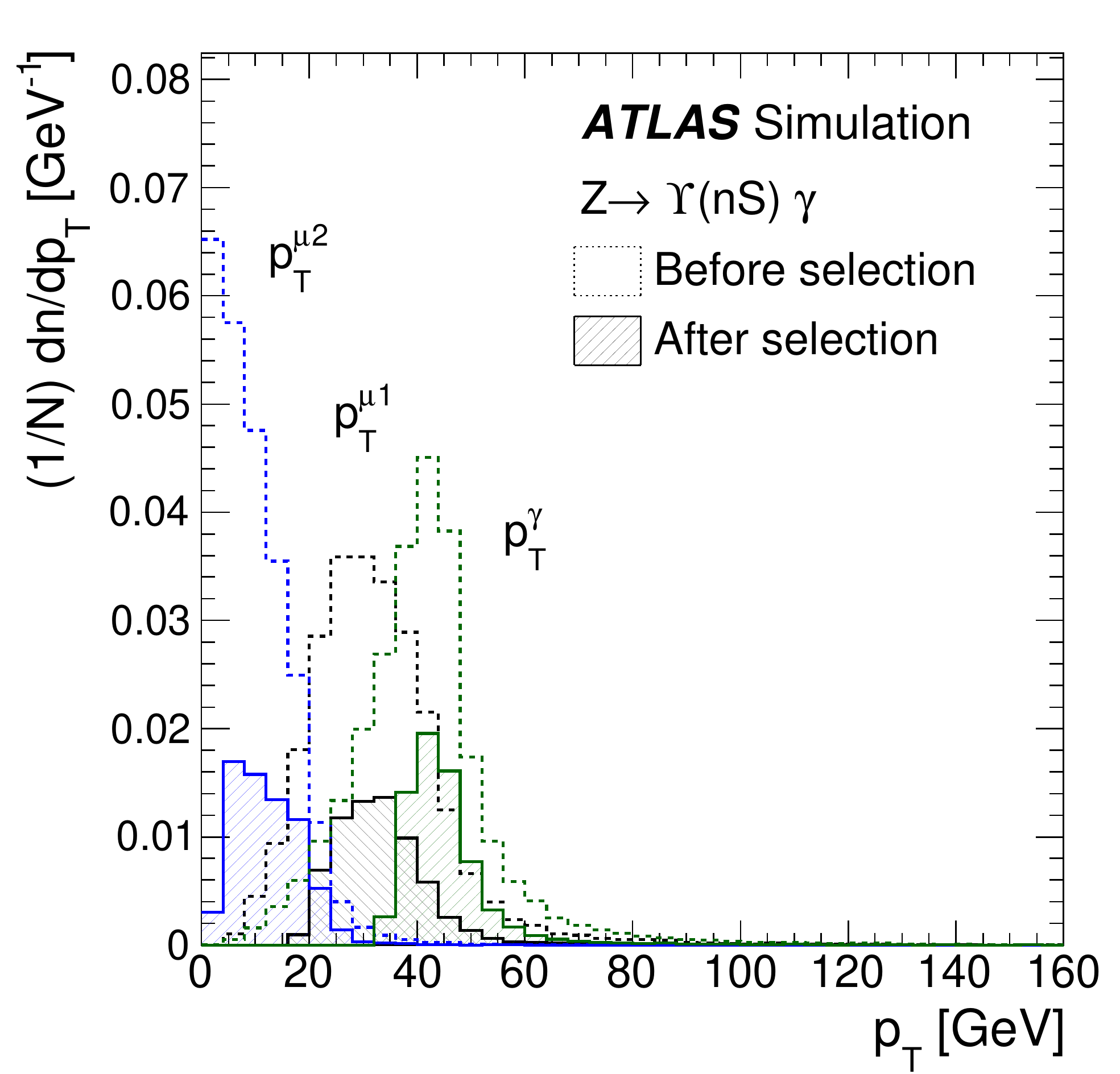}}
\caption{Generator-level transverse momentum ($\pt$) distributions of the photon and of the muons, ordered in $\pt$, for \subref{fig:pt_ph_muons_Hjpsi} $H\to\Jpsi\,\gamma$, \subref{fig:pt_ph_muons_Zjpsi} $Z\to\Jpsi\,\gamma$, \subref{fig:pt_ph_muons_Hupsi} $H\to\Upsilon(nS)\,\gamma$ and \subref{fig:pt_ph_muons_Zupsi} $Z\to\Upsilon(nS)\,\gamma$ simulated signal events, respectively. The leading muon candidate is denoted by $\pt^{\mu 1}$ and the subleading candidate by $\pt^{\mu 2}$. The hatched histograms denote the full event selection while the dashed histograms show the events at generator level that fall within the analysis geometric acceptance (both muons are required to have $|\etamu|<2.5$ while the photon is required to have $|\etagamma|<2.37$, excluding the region $1.37<|\etagamma|<1.52$). The dashed histograms are normalised to unity, and the relative difference between the two sets of distributions corresponds to the effects of reconstruction, trigger, and event selection efficiencies. \label{fig:pt_ph_muons}}
\end{figure}

The $m_{\mumu\,\gamma}$ resolution is 1.6--1.8\% for both the Higgs
and $Z$~boson decays. For each of the final states, a two-dimensional ($m_{\mumu\gamma}$ and $m_{\mumu}$) probability density function (pdf) is used to model the signal. The Higgs boson signals are modelled with two-dimensional multivariate Gaussian distributions, which retain the correlation between $m_{\mumu\gamma}$ and $m_{\mumu}$ of the final states. For the $Z$ boson decays, the $m_{\mumu\gamma}$ distributions of the signal are modelled with Voigtian pdfs (a convolution of Breit--Wigner and Gaussian pdfs) corrected with mass-dependent efficiency factors, and the $m_{\mumu}$ distributions are modelled as Gaussian pdfs.

The $m_{\mumu}$ distribution for selected $\psi(nS)\,\gamma$ candidates, which pass the background generation region criteria described in Section \ref{sec:backgroundmodel}, is shown in Figure~\ref{fig:MassJpsiFitINC} and
exhibits clear peaks at the $\Jpsi$ and $\psi(2S)$ masses. In Figures~\ref{fig:MassUpsilonFitB} and  \ref{fig:MassUpsilonFitEC}, the corresponding distributions for the
selected $\Upsilon(nS)\,\gamma$ candidates are shown, where the
$\Upsilon(1S,2S,3S)$ peaks can be observed. The $\psi(nS)$ and $\Upsilon(nS)$ peaks are fitted with Gaussian pdfs and are used to cross-check the parameters obtained from the fit to simulated signal event samples, while the background is modelled with a Chebychev polynomial function. The experimental resolution in
$m_{\mumu}$ is approximately $54~\MeV$ for the $\Jpsi\,\gamma$ candidates ($43~\MeV$ for events in the barrel category and $64~\MeV$ for events in the endcap category).

\begin{figure}[h]
\centering
\subfigure[\label{fig:MassJpsiFitINC}]{\includegraphics[width=0.45\linewidth]{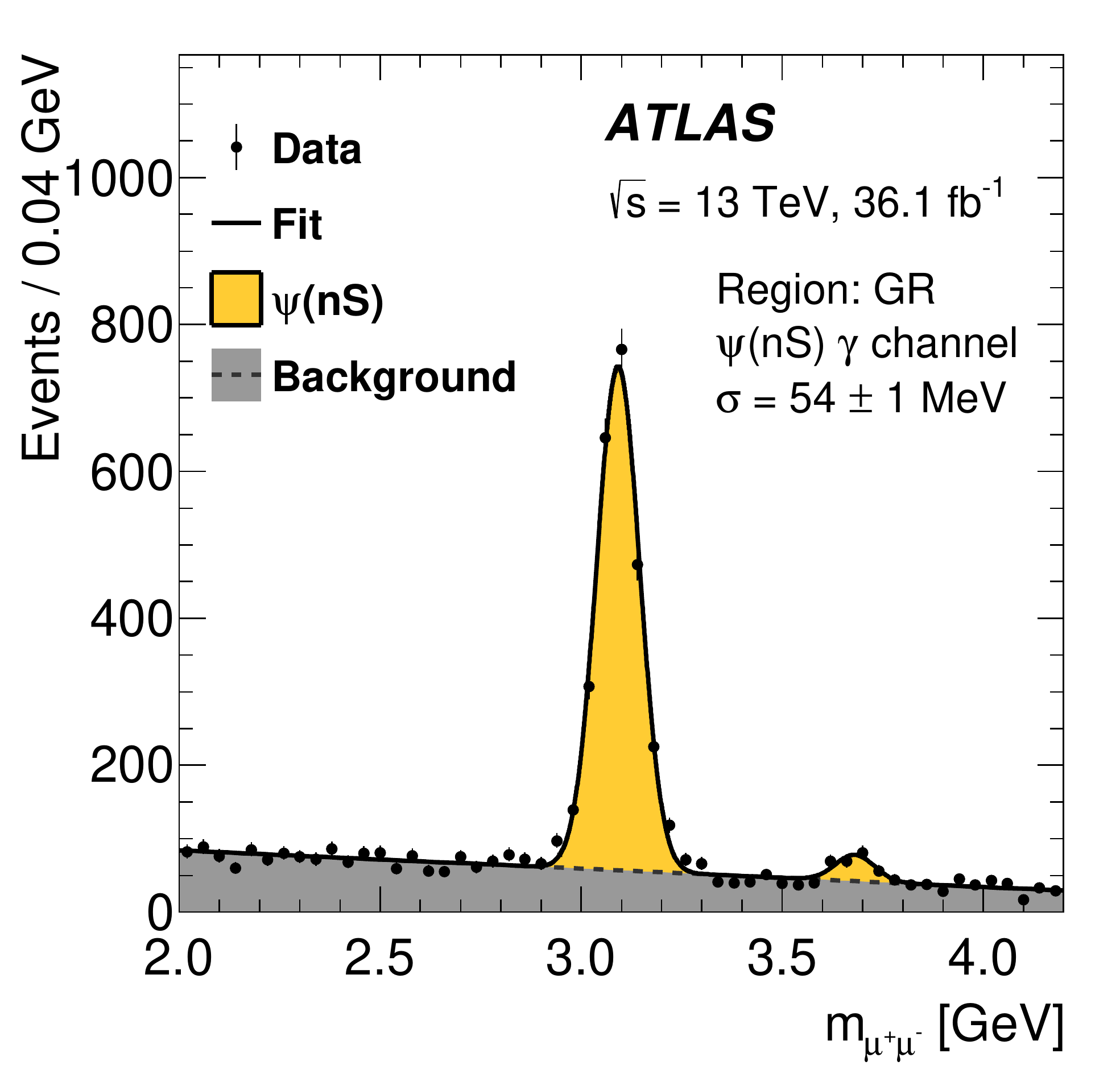}}

\subfigure[\label{fig:MassUpsilonFitB}]{\includegraphics[width=0.45\linewidth]{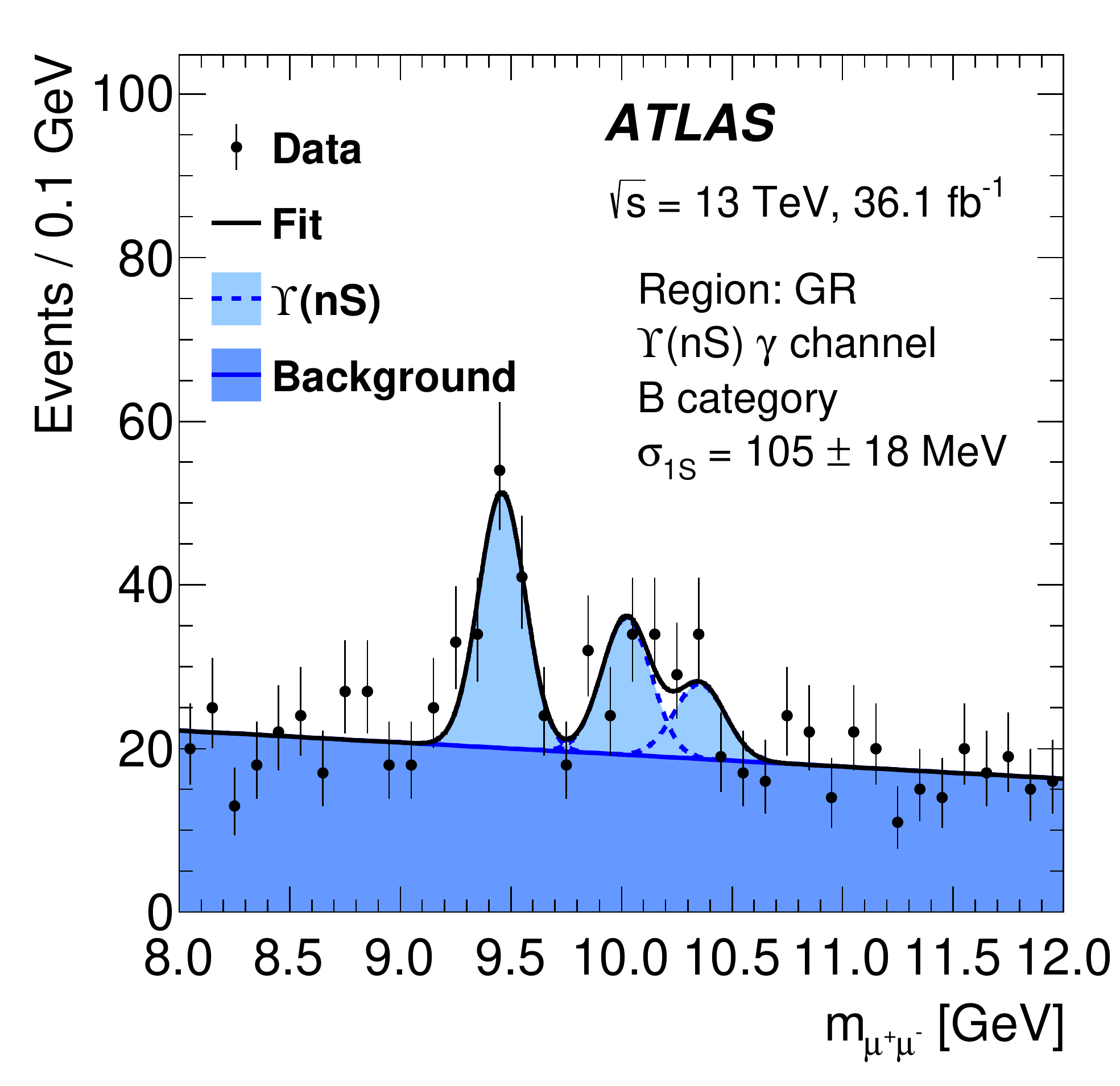}}
\subfigure[\label{fig:MassUpsilonFitEC}]{\includegraphics[width=0.45\linewidth]{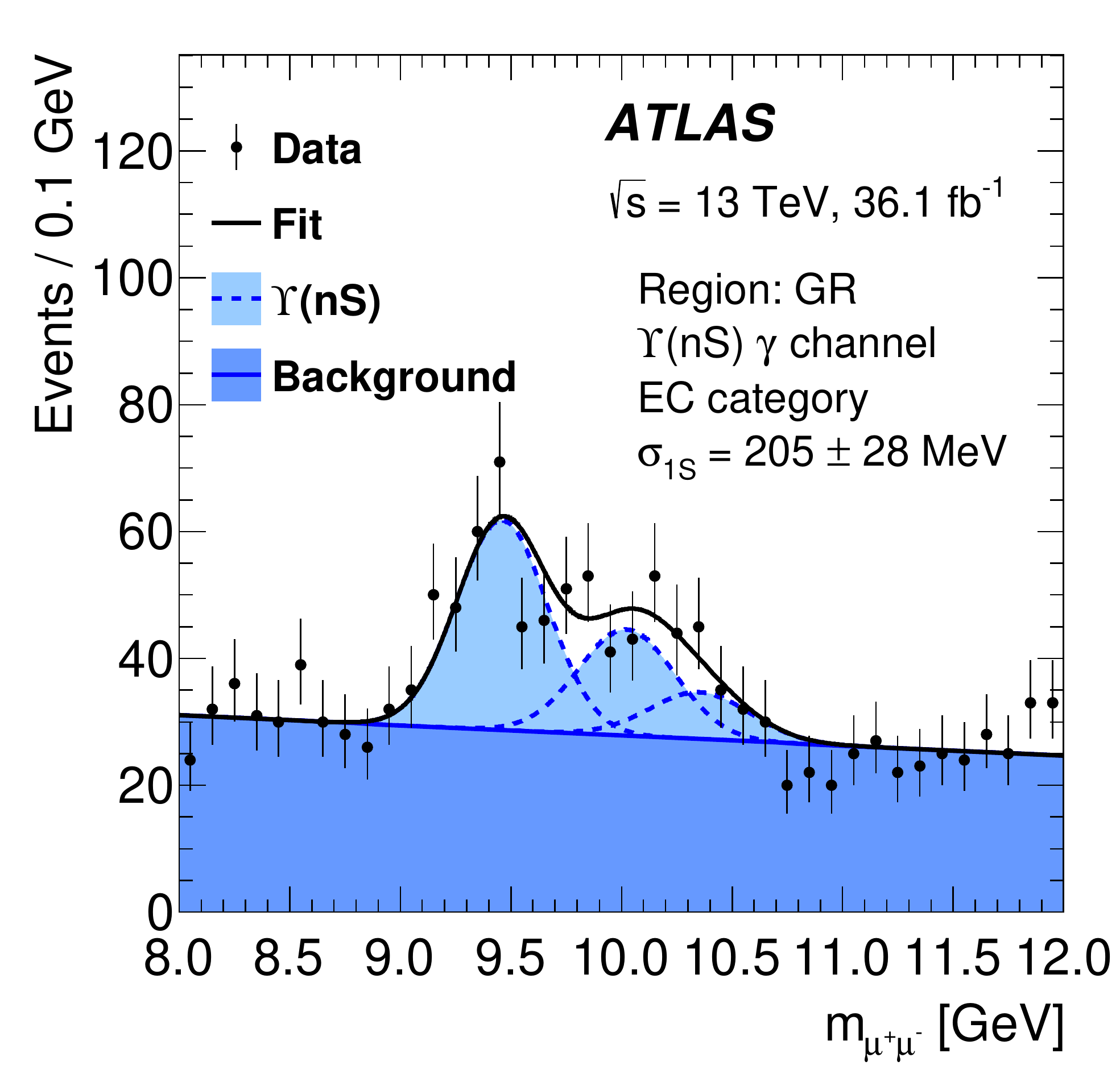}}
\vspace{-0.1cm}
\caption{Distribution of $\mumu$ invariant mass for \subref{fig:MassJpsiFitINC} $\psi(nS)\,\gamma$ and $\Upsilon(nS)\,\gamma$ (\subref{fig:MassUpsilonFitB} barrel and \subref{fig:MassUpsilonFitEC} endcap categories) candidates. The candidates satisfy the event selection but without the nominal isolation requirements and with a looser minimum $\pt^{\cal{Q}}$ requirement of $30~\GeV$. These events constitute the background ``generation region'' defined in Section~\ref{sec:backgroundmodel}.
\label{fig:MassFit}}
\end{figure}

\section{Background modelling}
\label{sec:backgroundmodel}

The dominant source of background exhibits a non-resonant $m_{\mumu\gamma}$ distribution and is
composed of two distinct contributions: genuine $\cal{Q}\to\mumu$ decays and non-resonant dimuon production.
The decay $Z\to\mumu\gamma$ with final-state radiation (\Zboson FSR)
constitutes a further, smaller background contribution exhibiting a characteristic resonant structure in the $m_{\mumu\gamma}$ distribution.

The $m_{\mumu\gamma}$ continuum background is modelled
with a non-parametric data-driven approach using templates to
describe the kinematic distributions. The normalisation
of the background is extracted directly from a fit to the data.
The shape of the background model in the final discriminant variable
is also profiled in the fit. A similar procedure was used in the
earlier search for Higgs and $Z$ boson decays into $\Jpsi\,\gamma$ and
$\Upsilon(nS)\,\gamma$~\cite{HIGG-2014-03} and the
searches for Higgs and $Z$ boson decays into
$\phi\,\gamma$ and $\rho\,\gamma$~\cite{HIGG-2016-05,HIGG-2016-13}.

The background model uses a sample of 5500
$\psi(nS)\,\gamma$ and 2300
$\Upsilon(nS)\,\gamma$ candidate events. These events pass all of the
kinematic selection requirements described previously, except that the
photon and $\cal{Q}$ candidates are not required to satisfy the
nominal isolation requirements and a looser minimum $\pt^{\cal{Q}}$ requirement of
$30~\GeV$ is imposed. These events define the background-dominated ``generation region''
(GR).
From these events, pdfs are constructed to describe the distributions
of the relevant kinematic and isolation variables and their most important correlations. The data control samples are corrected for contamination from $Z\to\mumu\gamma$ decays.

The pdfs of these kinematic and isolation variables are
sampled to generate an ensemble of pseudocandidates, each with
complete $\cal{Q}$ and $\gamma$ four-vectors and the associated
$\cal{Q}$ and photon isolation values. The important correlations
among the kinematic and isolation variables of the background events,
in particular between $\pt^{\cal Q}$ and $\ptgamma$,
are retained in the generation of the pseudocandidates through the
following sampling scheme:

\begin{itemize}
  \item Initially, a value for $\pt^{\cal{Q}}$ is sampled from the corresponding pdf.

  \item The distribution of $\ptgamma$ is parameterised in bins of $\pt^{\cal{Q}}$, and a value is sampled from the corresponding bin given the previously sampled value of $\pt^{\cal{Q}}$.
The muon isolation variables are parameterised in bins of $\pt^{\cal{Q}}$ and $\ptgamma$ and sampled according to the previously selected values.

\item The distributions of the pseudorapidity difference between $\cal{Q}$ and $\gamma$, $\Delta\eta(\cal{Q},\gamma)$, the photon calorimeter isolation variable, and their correlations are parameterised in a binned two-dimensional distribution in the same bins of $\pt^{\cal{Q}}$ used to describe the $\ptgamma$ and muon isolation variables.

\item  Given the selected values of relative photon calorimeter isolation and $\pt^{\cal{Q}}$, a value for the relative photon track isolation is sampled.

\item Given the selected values of $\Delta\eta(\cal{Q},\gamma)$ and $\pt^{\cal{Q}}$, a value is sampled for the azimuthal angular separation between $\cal{Q}$ and $\gamma$, $\Delta\phi(\cal{Q},\gamma)$.

\item Values for $\eta^{\cal{Q}}$ and $\phi^{\cal{Q}}$ are sampled from a binned histogram of the corresponding distributions in the data control sample. These are combined with $\Delta\eta(\cal{Q},\gamma)$ and $\Delta\phi(\cal{Q},\gamma)$ to give the values of $\etagamma$ and $\phi^{\gamma}$.

\item A value for $m_{\cal{Q}}$ is sampled from within the required region of $m_\mumu$. Separate pdfs are used to describe the $m_\mumu$ distributions of resonant $\psi(nS)$ and $\Upsilon(nS)$ production and non-resonant dimuons, which is referred to as ``combinatoric'' in the following.

\end{itemize}

The use of this procedure ensures a good description of the background and avoids any reliance on the accuracy or limited sample size of simulated background events.

The nominal selection requirements are imposed and the
surviving pseudocandidates are used to construct templates for the
$m_{\cal{Q}\,\gamma}$ distributions which are then smoothed using a Gaussian kernel density
estimation~\cite{Cranmer:2000du}.
Potential contamination of the GR sample from signal events is
expected to be negligible, and it was verified, through signal
injection tests, that such a potential signal contamination would not
affect the shape of the background model.

The normalisation of the exclusive background from $Z$ FSR is determined directly from the fit to the data. The shape of this background in the $m_{\mumu\gamma}$ distribution is modelled with a Voigtian pdf, while the $m_\mumu$ distribution is modelled with a first order polynomial. The parameters of the former pdf are derived from the simulated $Z\to\cal{Q}\,\gamma$ signal samples. The parameters of the latter pdf are determined directly in the fit to data.

To validate this background model with data, the $m_{\cal{Q}\,\gamma}$
distributions in several regions defined by kinematic and
isolation requirements looser than the nominal signal requirements
are used to compare the prediction of the background model with the
data.
Three validation regions are defined using the GR selection as the basis
and adding either the $\pt^{\cal{Q}}$ requirement (VR1), or the muon isolation requirements (VR2),
or the photon isolation requirement (VR3).
The $m_{\cal{Q}\,\gamma}$ distributions in these validation regions are shown in Figure~\ref{fig:BkgdVal} for the $\psi(nS)\,\gamma$ and the $\Upsilon(nS)\,\gamma$ final states. The background model is found to describe the data well in all validation regions.

\begin{figure}[!htbp]
  \begin{center}
    \hspace{-0.50cm}
    \subfigure[]{\label{fig:BkgdValJpsi_VR1}\includegraphics[width=0.34\textwidth]{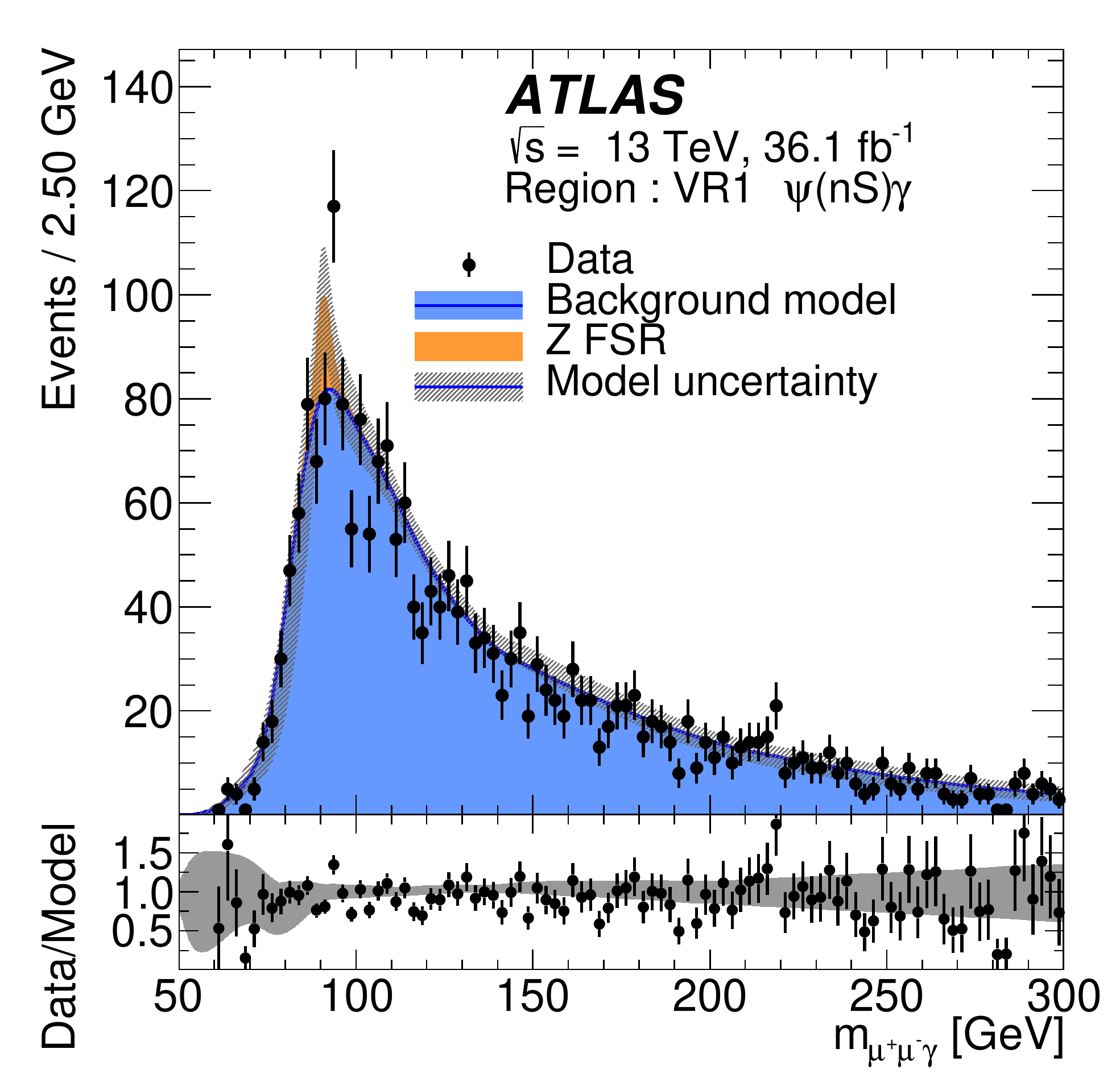}} \hspace{-0.30cm}
    \subfigure[]{\label{fig:BkgdValJpsi_VR2}\includegraphics[width=0.34\textwidth]{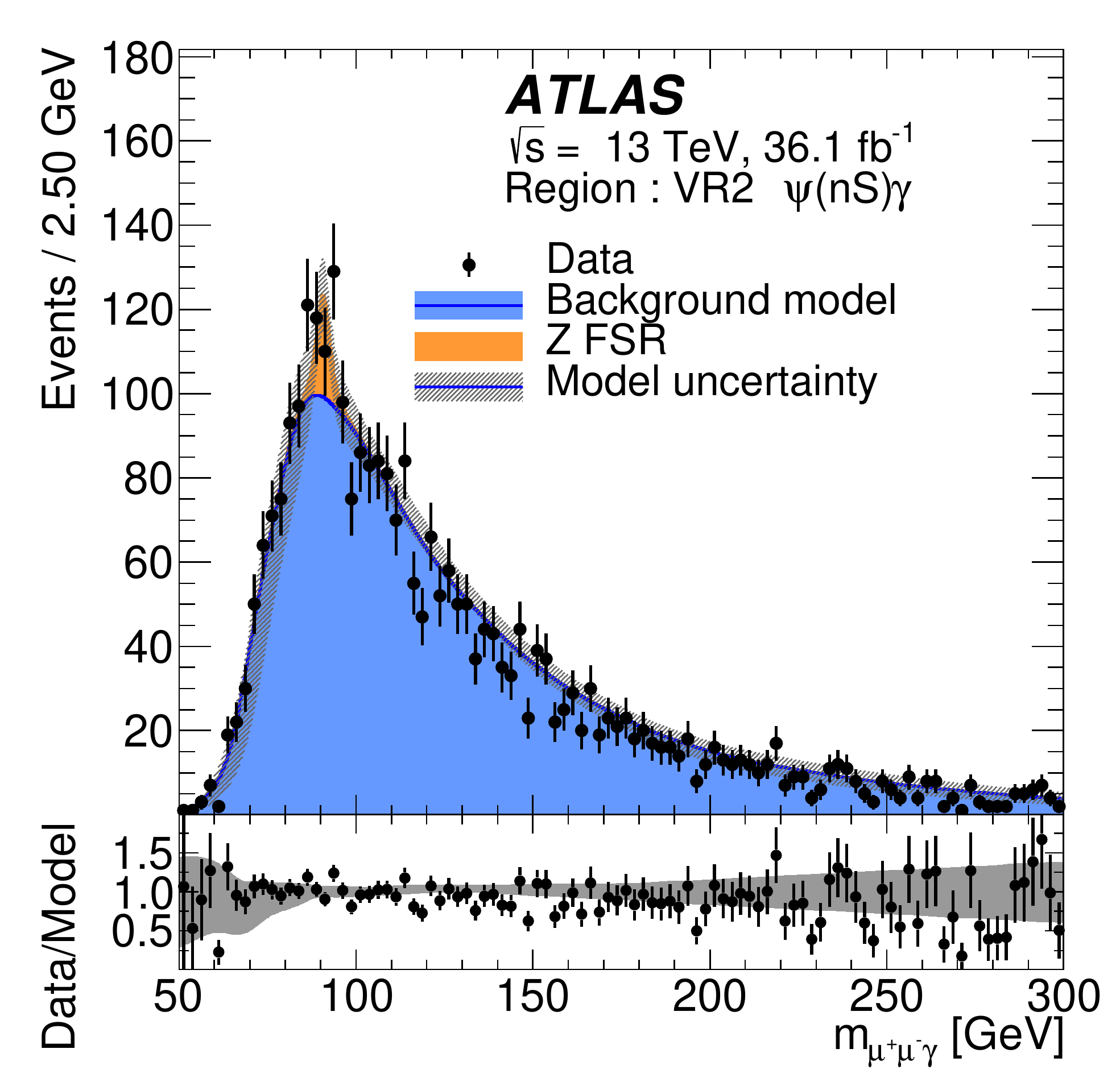}} \hspace{-0.30cm}
    \subfigure[]{\label{fig:BkgdValJpsi_VR3}\includegraphics[width=0.34\textwidth]{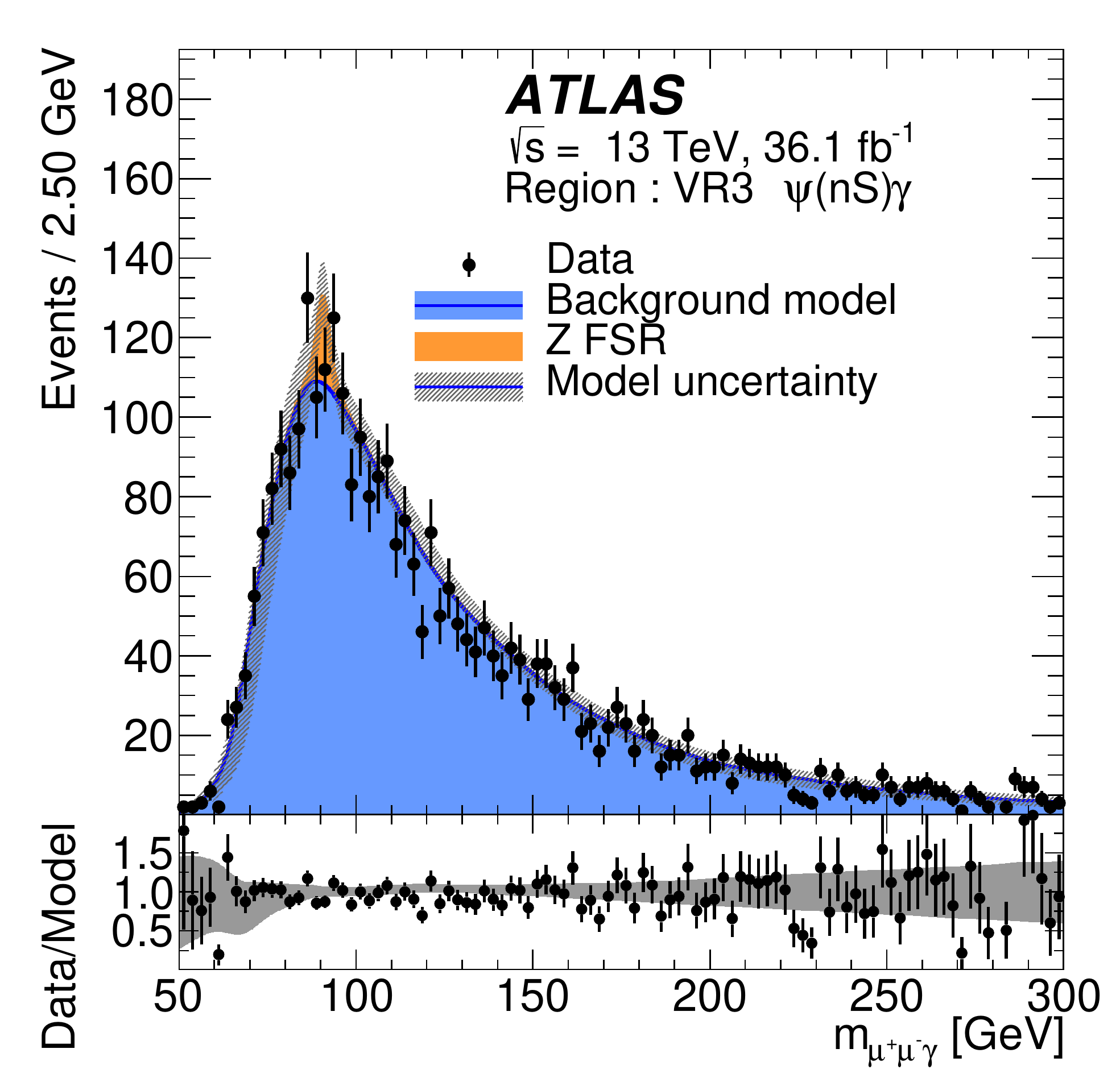}}

    \hspace{-0.50cm}
    \subfigure[]{\label{fig:BkgdValUpsilon_VR1}\includegraphics[width=0.34\textwidth]{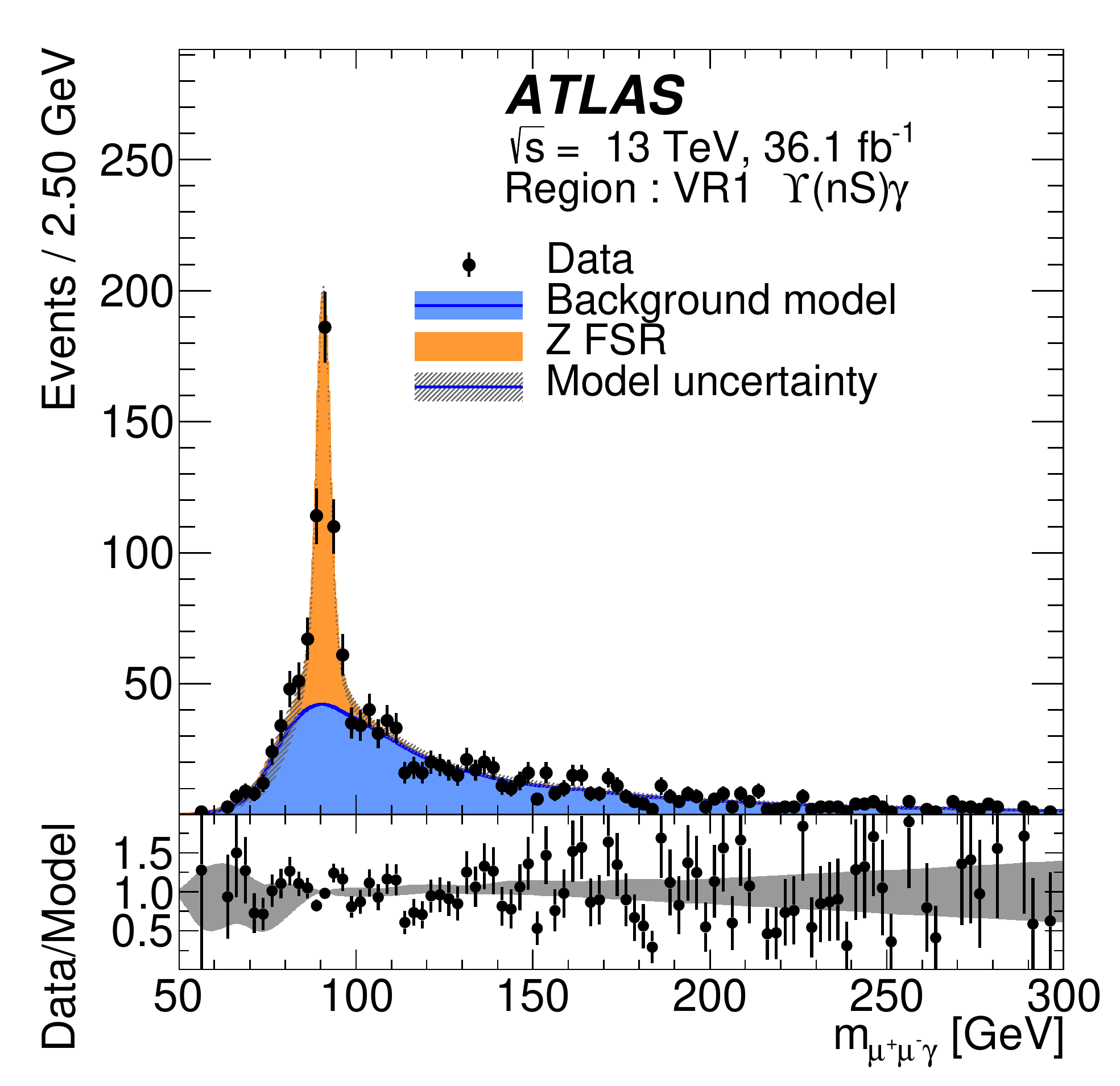}} \hspace{-0.30cm}
    \subfigure[]{\label{fig:BkgdValUpsilon_VR2}\includegraphics[width=0.34\textwidth]{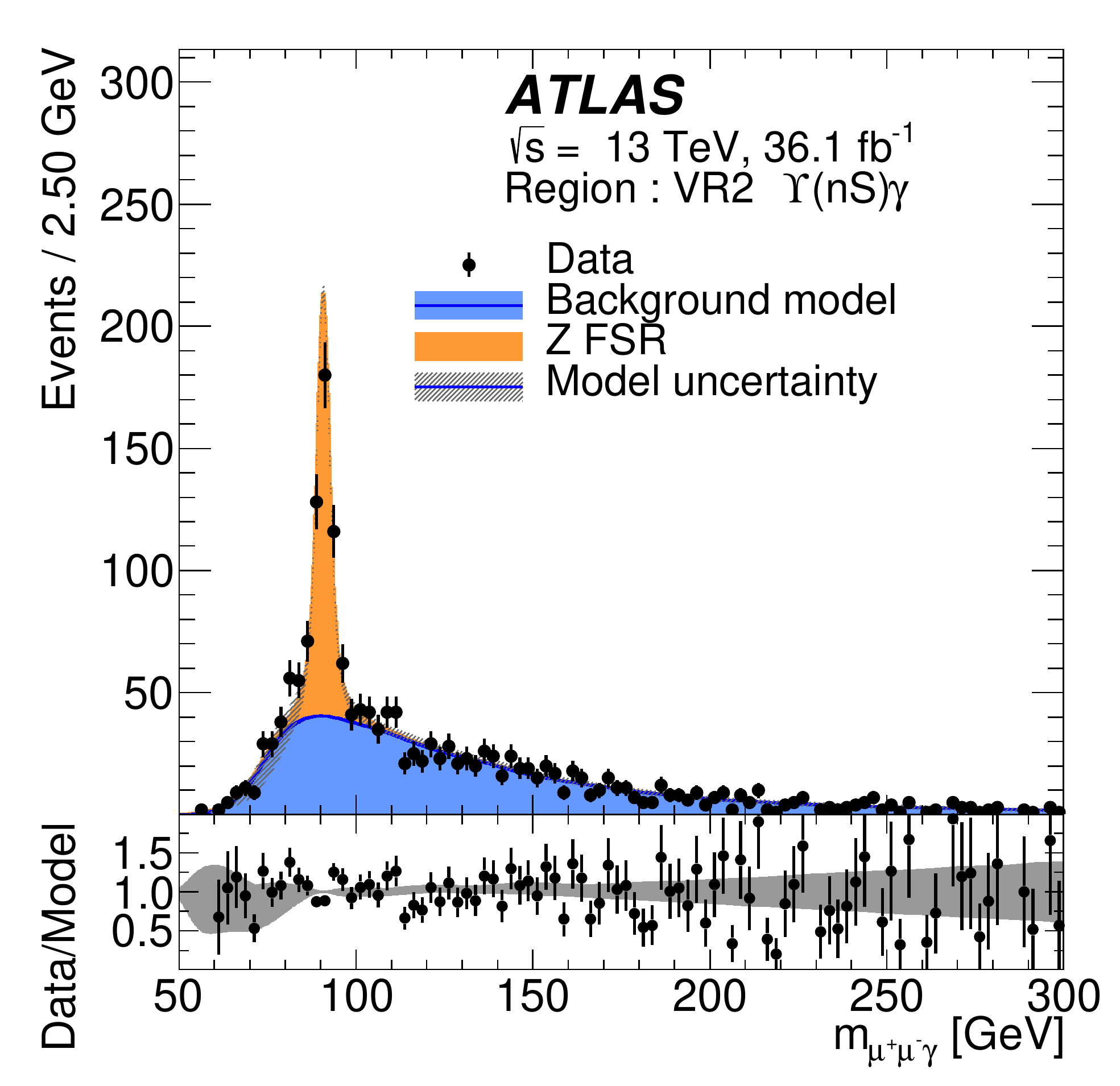}} \hspace{-0.30cm}
    \subfigure[]{\label{fig:BkgdValUpsilon_VR3}\includegraphics[width=0.34\textwidth]{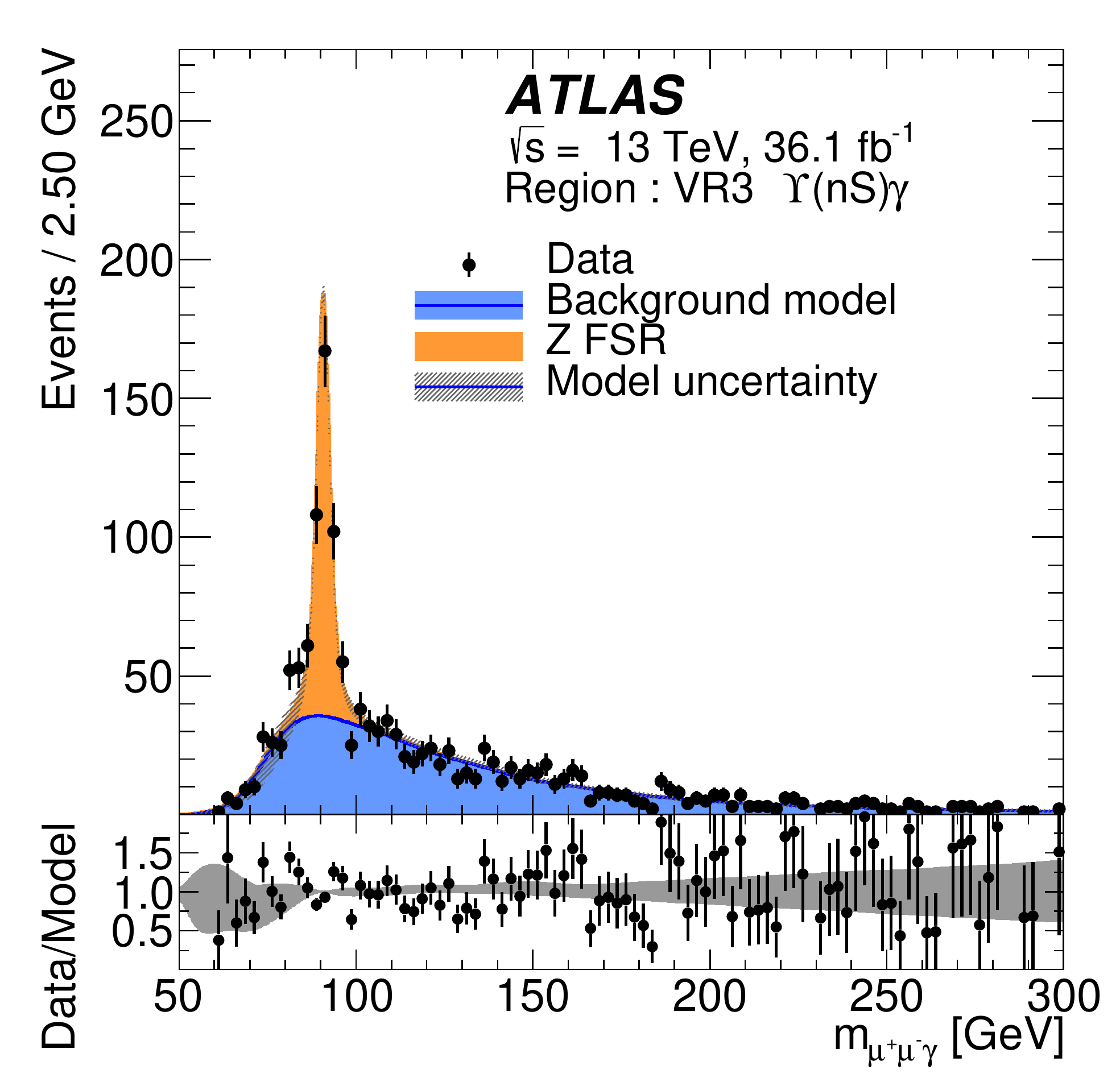}}
  \end{center}
\vspace{-0.6cm}
\caption{The distribution of $m_{\mumu\gamma}$ in data compared to the prediction of the background model for (\subref{fig:BkgdValJpsi_VR1}, \subref{fig:BkgdValJpsi_VR2} and \subref{fig:BkgdValJpsi_VR3}) $H(Z)\to\psi(nS)\,\gamma$ and (\subref{fig:BkgdValUpsilon_VR1}, \subref{fig:BkgdValUpsilon_VR2} and \subref{fig:BkgdValUpsilon_VR3}) $H(Z)\to\Upsilon(nS)\,\gamma$ in the VR1, VR2 and VR3 validation regions. $Z$ FSR refers to the $Z\to\mumu\gamma\,$ background contribution. The background model is normalised to the observed number of events within the region shown. The uncertainty band corresponds to the uncertainty envelope
derived from variations in the background modelling procedure.\label{fig:BkgdVal}
}
\end{figure}

\section{Systematic uncertainties}
\label{sec:systematic}

The systematic uncertainties in the expected signal yields are
summarised in Table~\ref{tab:systematics}. Uncertainties in the Higgs production cross sections are obtained from Ref.~\cite{deFlorian:2016spz}. The $Z$ boson production cross section uncertainty is taken from the measurement in Ref.~\cite{STDM-2015-03}.

The uncertainty in the acceptance of the Higgs boson signal
due to missing higher orders in the generator, parton distribution
functions, underlying-event set of tuned parameters, and parton showering is estimated by
studying the effect of the different variations on the acceptance at
generator level. The total uncertainty in the acceptance due to these
effects is estimated to be 1.8\%.
For the $Z$ boson, the respective uncertainty is determined to be 6\% by comparing the generator-level acceptance predictions of the nominal simulated sample with those of \progname{Madgraph5\_aMC@NLO} v2.2.2~\cite{Alwall:2014hca} and \progname{Sherpa} 2.2.1~\cite{Gleisberg:2008ta}.

Trigger efficiencies for photons are determined
from samples enriched with $Z\to\ee$ events in
data~\cite{ATLAS-CONF-2011-114}. The systematic
uncertainty in the expected signal yield associated with the trigger
efficiency is estimated to be 2.0\%. Photon identification efficiencies are determined using the enriched $Z\to\ee$ events, as well as inclusive photon events and $Z\to\ell\ell \gamma$ events~\cite{Aaboud:2016yuq,ATL-PHYS-PUB-2016-014}. The impact on the yields of the photon identification
efficiency uncertainties, for both the converted and unconverted
photons, are 1.4\% for the Higgs and
$Z$~boson signals. The effect of the muon reconstruction and identification efficiency uncertainty
is 2.8\%~\cite{PERF-2015-10}.

The photon energy scale uncertainty, determined from $Z\to\ee$ events
and validated using $Z\to\ell\ell \gamma$
events~\cite{ref:run1-egamma-calib,ATL-PHYS-PUB-2016-015}, is propagated through the
simulated signal samples as a function of $\etagamma$ and $\ptgamma$.
The uncertainty associated with the determination of the photon energy
scale and resolution in the simulation is 0.3\% in the yield of the Higgs and $Z$ boson signal. Similarly, the systematic uncertainty associated
with the scale of the muon momentum measurement is 0.2\%~\cite{PERF-2015-10}. The effect of the energy/momentum scale and resolution uncertainties on the signal shape is negligible for the Higgs and $Z$ boson signals. For the $Z$ FSR background shape, the effect is 0.2\% on the mean value of the $m_{\mumu\gamma}$ distribution. Including the corresponding nuisance parameter results in a 0.5\% change in the expected limit of the $Z\to\Upsilon(nS)\gamma$ channels with negligible impact in the other channels.

The shape of the background model is allowed to vary around the
nominal shape, and the parameters controlling these systematic variations
are treated as nuisance parameters in the maximum-likelihood fit described in Section \ref{sec:results}.
Three such shape variations are implemented through
scale variations on the $\ptgamma$ distribution in the model, linear
distortions of the shape of the $\Delta\phi({\cal{Q}},\gamma)$ distribution,
and a global tilt of the three-body mass distribution around a pivot
point.
The first two variations are straightforward alterations to the
underlying kinematics of the pseudocandidates, which cause
corresponding changes in the three-body mass. The third variation is
applied directly to the final three-body mass template.
These systematic uncertainties allow the background template to
adjust to the observed distribution in data.

\begin{table}[h]
  \centering
  \caption{Summary of the systematic uncertainties in the expected signal yields.\label{tab:systematics}}
\begin{tabular}{lcc}
\hline
\multirow{2}{*}{Source of systematic uncertainty} & \multicolumn{2}{c}{Yield uncertainty} \\
& $H(Z)\to{\cal{Q}}\,\gamma$ \\ \hline
Total $H$($Z$) cross section & {$7.0\%$ ($2.9\%$)} \\
Integrated luminosity & $2.1\%$ \\
$H$($Z$) QCD modelling & {$1.8\%$ ($6\%$)} \\
Trigger efficiency  & $2.0\%$ \\
Photon identification & $1.4\%$ \\
Muon identification and reconstruction  & $2.8\%$ \\
Photon energy scale & $0.3\%$ \\
Muon momentum scale & $0.2\%$ \\ \hline
\end{tabular}
\end{table}

\section{Results}
\label{sec:results}

The data are compared with background and signal predictions using a two-dimensional (2D) simultaneous
unbinned maximum-likelihood fit to the $m_{\mumu\,\gamma}$ and $m_{\mumu}$ distributions. A simultaneous 2D fit is required to distinguish the $Z$ FSR background from the $Z\to{\cal Q}\gamma$ signal and the non-resonant background. The parameters of interest are the Higgs and $Z$ boson
signal normalisations. Systematic uncertainties are modelled using
additional nuisance parameters in the fit; in particular, the
background normalisations are free parameters.
The fit uses the selected events with $m_{{\cal Q}\gamma}<300~\GeV$.

In total, 1033 events were observed in the $\psi(nS)\,\gamma$ and 906
in the $\Upsilon(nS)\,\gamma$ signal regions. The
expected and observed numbers of background events within the
$m_{\cal{Q}\,\gamma}$ ranges relevant to the Higgs and $Z$ boson
signals are shown in Table~\ref{tab:events}.
The results of the background-only fits for the $\psi(nS)\,\gamma$ and $\Upsilon(nS)\,\gamma$ analyses are shown in
Figure~\ref{fig:Fit}.

\begin{table}[h]
  \centering
  \caption{The number of observed events and the mean expected background, with its total uncertainty,
  for the $m_{\cal{Q}\,\gamma}$ ranges of interest. The expected $Z$ and Higgs boson contributions are shown for branching fraction
  values of $10^{-6}$ and $10^{-3}$, respectively. These values are motivated by the expected sensitivity of the search to the respective branching fractions.
  \label{tab:events}}
  \begin{tabular}{|lc|rr|rr|rr|c}
\hline
\multicolumn{2}{|c|}{\multirow{3}{*}{$m_{\mumu}$ mass range [\GeV]}} & \multicolumn{4}{c|}{Observed (expected background)}&\multicolumn{1}{c|}{$Z$ signal} & \multicolumn{1}{c|}{$H$ signal}\\ \cline{3-6}
& & \multicolumn{4}{c|}{$m_{\mumu\gamma}$ mass range [\GeV]} & \multicolumn{1}{c|}{for} & \multicolumn{1}{c|}{for} \\ \cline{3-6}
& & \multicolumn{2}{c|}{$81$--$101$} & \multicolumn{2}{c|}{$120$--$130$}  & \multicolumn{1}{c|}{${\cal B}=10^{-6}$} & \multicolumn{1}{c|}{${\cal B}=10^{-3}$} \\ \hline
$\Jpsi\,\gamma$ & \hphantom{0}2.9--3.3\hphantom{0}  & 92  & (89 $\pm$ 6) & 20 & (23.6 $\pm$ 1.3)  & 13.7 $\pm$ 1.1\hphantom{0} & 22.2 $\pm$ 1.9\hphantom{0}  \\
$\psi(2S)\,\gamma$ & \hphantom{0}3.5--3.9\hphantom{0}  & 43 & (42 $\pm$ 5) & 8 & (10.0 $\pm$ 0.8)  & 1.82 $\pm$ 0.14 & 2.96 $\pm$ 0.25\\
$\Upsilon(1S)\,\gamma$ & \hphantom{0}9.0--10.0  & 115 & (126 $\pm$ 8)  & 9 & (13.6 $\pm$ 1.2)  &\hphantom{0}7.8 $\pm$ 0.6\hphantom{0} & 10.7 $\pm$ 0.9\hphantom{0}  \\
$\Upsilon(2S)\,\gamma$ & \hphantom{0}9.5--10.5  & 106 & (121 $\pm$ 8)  & 8 & (12.6 $\pm$ 1.4)  & \hphantom{0}5.9 $\pm$ 0.5\hphantom{0} & \hphantom{0}8.1 $\pm$ 0.7\hphantom{0}\\
$\Upsilon(3S)\,\gamma$ & 10.0--11.0  & 112 & (113 $\pm$ 8) & 7 & (10.6 $\pm$ 1.2)  & \hphantom{0}7.1 $\pm$ 0.6\hphantom{0} & \hphantom{0}9.2 $\pm$ 0.8\hphantom{0}\\
\hline
\end{tabular}
\end{table}

\begin{figure}[p]
  \begin{center}
    \subfigure[\label{fig:FitJpsi}]{\includegraphics[width=0.8\linewidth]{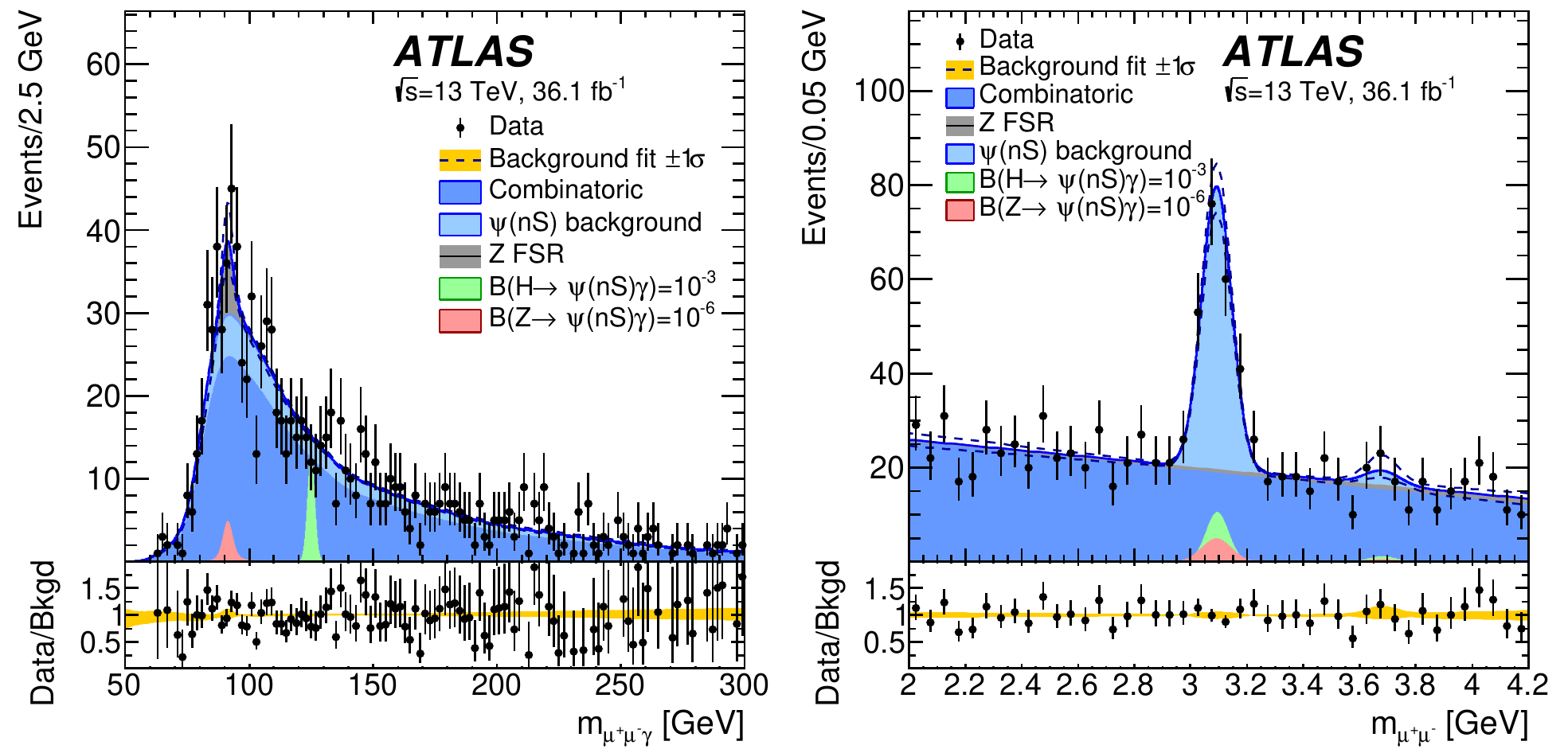}}

    \subfigure[\label{fig:FitUpsiB}]{\includegraphics[width=0.8\linewidth]{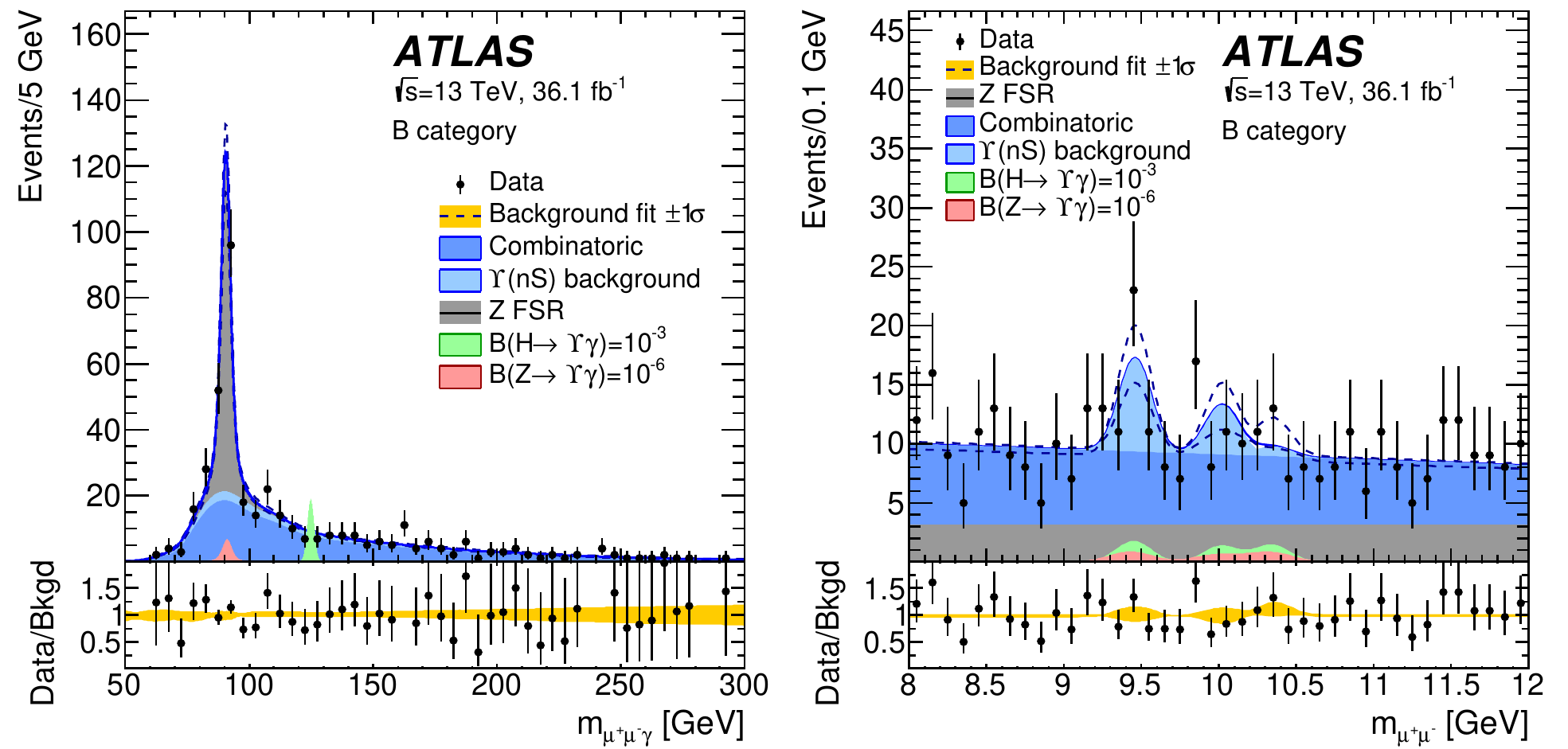}}

    \subfigure[\label{fig:FitUpsiEC}]{\includegraphics[width=0.8\linewidth]{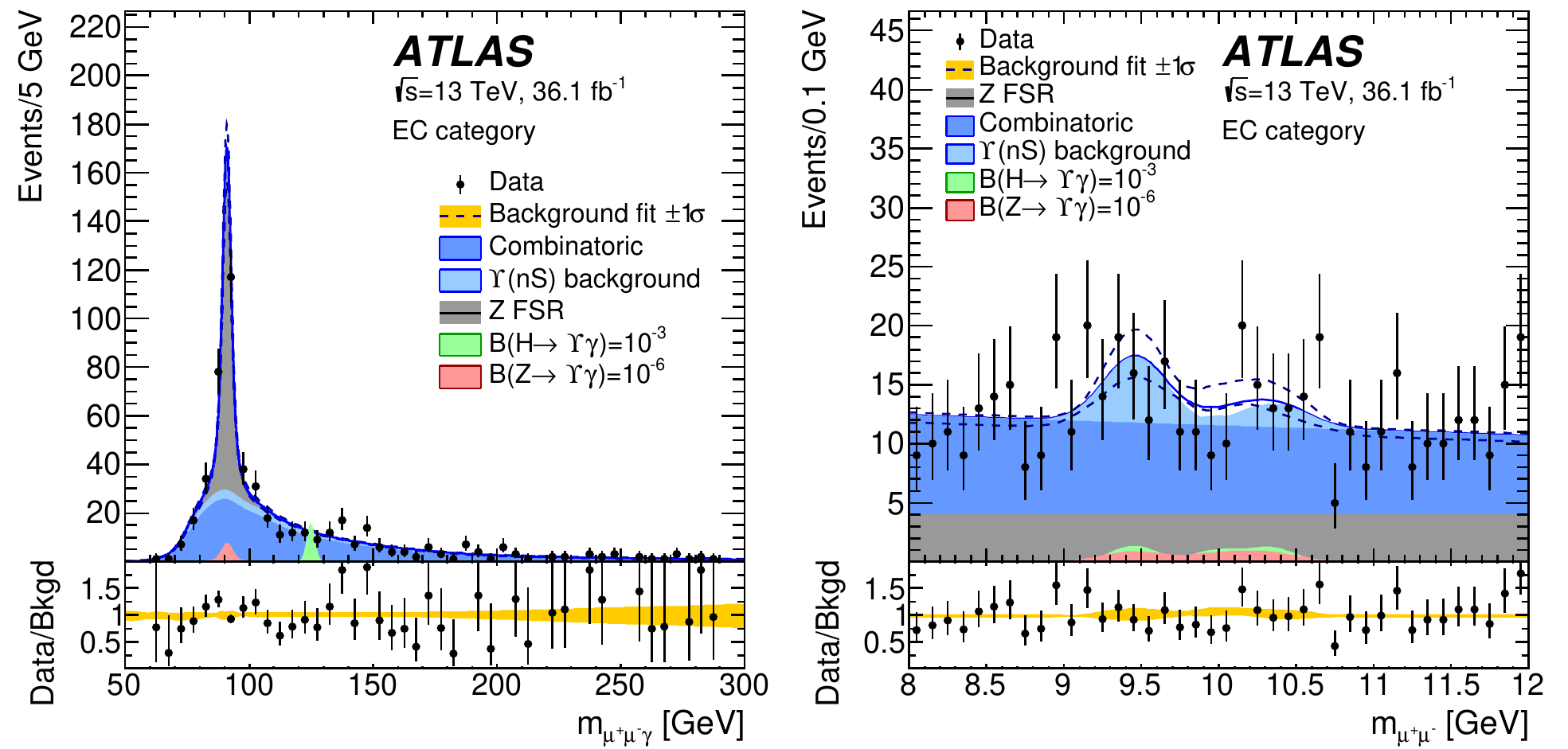}}
  \end{center}
  \vspace{-0.6cm}
  \caption{The $m_{\mumu\,\gamma}$ and $m_{\mumu}$ distributions for the selected \subref{fig:FitJpsi} $\psi(nS)\,\gamma$ and $\Upsilon(nS)\,\gamma$ (\subref{fig:FitUpsiB} barrel and \subref{fig:FitUpsiEC} endcap categories) candidates along with the results of the maximum-likelihood fits with background-only models. \texttt{Z FSR} refers to the $Z\to\mumu\gamma\,$ background contribution. The solid blue line denotes the full fit result and the dashed blue lines correspond to its $\pm1\sigma$ uncertainty band. The ratios of the data to the background-only fits are also shown. \label{fig:Fit}}
\end{figure}

The systematic uncertainties described in Section~\ref{sec:systematic}
result in a 1.0\% increase of the expected 95\%~CL upper limit on the
branching fraction of the $H\to\psi(nS)\,\gamma$ decays. For the $Z\to\psi(nS)\,\gamma$
decays, the effect is larger, 2.6\%, mostly due to the systematic
uncertainty in the background shape. Similar behaviour is observed in the $\Upsilon(nS)\,\gamma$ analysis with
systematic uncertainties resulting in a 2.5--2.7\% deterioration in the
sensitivity to the $H\to\Upsilon(nS)\,\gamma$ decays and a 2.8--2.9\% deterioration
in the sensitivity to the $Z\to\Upsilon(nS)\,\gamma$ decays, also mostly due to the systematic
uncertainty in the background shape.

On the basis of the fit to the observed data, the largest excess observed is 2.2$\sigma$ in the search for $Z\to\Jpsi\,\gamma$.
Upper limits are set on the
branching fractions for the Higgs and $Z$~boson decays into
$\cal{Q}\,\gamma$ using the CL$_\mathrm{s}$ modified frequentist
formalism~\cite{cls} with the profile-likelihood-ratio test
statistic~\cite{10.2307/91247} and the asymptotic approximations derived in Ref.~\cite{Cowan:2010js}. The expected SM production cross
section is assumed for the Higgs boson~\cite{deFlorian:2016spz}, while
the ATLAS measurement of the inclusive \Zboson~boson cross section is used
for the $Z$~boson signal~\cite{STDM-2015-03}, as discussed in
Section~\ref{sec:datamc}. The results are summarised in
Table~\ref{fig:Limit}. The observed 95\% CL upper limits on the branching fractions for Higgs and $Z$ boson decays into $J/\psi\,\gamma$ and $\psi(2S)\,\gamma$ are $(3.5,20)\times 10^{-4}$ and $(2.3,4.5)\times 10^{-6}$, respectively. The corresponding limits for the
Higgs and $Z$ boson decays into $\Upsilon(nS)\,\gamma$ ($n=1,2,3$) are $(4.9,5.9,5.7)\times 10^{-4}$ and
$(2.8,1.7,4.8)\times 10^{-6}$, respectively. Upper limits at 95\% CL on the product of the production cross section times branching fraction are determined for the Higgs boson decays, yielding
19~${\textrm{fb}}$~for the $H\to\Jpsi\,\gamma$ decay, 110~${\textrm{fb}}$~for the $H\to\psi(2S)\,\gamma$ decay, and (28, 33, 32)~${\textrm{fb}}$~for the $H\to\Upsilon(nS)\,\gamma$ ($n=1,2,3$) decays.
\begin{table}[h!]
\centering
\caption{Expected and observed branching fraction upper limits at 95\% CL
for the $H(Z)\to\,\Jpsi\,\gamma$, $H(Z)\to\,\psi(2S)\,\gamma$, and $H(Z)\to\Upsilon(nS)\,\gamma$ ($n=1,2,3$) analyses, assuming SM production for the Higgs and \Zboson\ bosons.
The $\pm 1\sigma$ intervals of the expected limits are also given.\label{fig:Limit}}
\begin{tabular}{l|r|r}
\hline
Branching fraction limit (95\% CL) & Expected & Observed \\ \hline
${\cal B}\left(H\to J/\psi\,\gamma\right)[\tabscript{10}{-4}{}]$ & $3.0_{-0.8}^{+1.4}$ & 3.5 \\
${\cal B}\left(H\to \psi\left(2S\right)\,\gamma\right)[\tabscript{10}{-4}{}]$ & $15.6_{-4.4}^{+7.7}$ & 19.8\\
${\cal B}\left(Z\to J/\psi\,\gamma\right)[\tabscript{10}{-6}{}]$ & $1.1_{-0.3}^{+0.5}$ & 2.3\\
${\cal B}\left(Z\to \psi\left(2S\right)\,\gamma\right)[\tabscript{10}{-6}{}]$ & $6.0_{-1.7}^{+2.7}$ & 4.5\\

${\cal B}\left(H\to\Upsilon(1S)\,\gamma\right)[\tabscript{10}{-4}{}]$ & $5.0_{-1.4}^{+2.4}$ & 4.9\\
${\cal B}\left(H\to\Upsilon(2S)\,\gamma\right)[\tabscript{10}{-4}{}]$ & $6.2_{-1.7}^{+3.0}$ & 5.9\\
${\cal B}\left(H\to\Upsilon(3S)\,\gamma\right)[\tabscript{10}{-4}{}]$ & $5.0_{-1.4}^{+2.5}$ & 5.7\\

${\cal B}\left(Z\to\Upsilon(1S)\,\gamma\right)[\tabscript{10}{-6}{}]$ & $2.8_{-0.8}^{+1.2}$ & 2.8 \\
${\cal B}\left(Z\to\Upsilon(2S)\,\gamma\right)[\tabscript{10}{-6}{}]$ & $3.8_{-1.1}^{+1.6}$ & 1.7 \\
${\cal B}\left(Z\to\Upsilon(3S)\,\gamma\right)[\tabscript{10}{-6}{}]$ & $3.0_{-0.8}^{+1.3}$ & 4.8 \\
\hline
\end{tabular}
\end{table}

\FloatBarrier
\section{Summary}
\label{sec:summary}
Searches for the exclusive decays of Higgs and $Z$ bosons into $\Jpsi\,\gamma$, $\psi(2S)\,\gamma$, and
$\Upsilon(nS)\,\gamma$ have been performed with a $\rts=13~\TeV$ $pp$ collision
data sample collected with the ATLAS detector at the LHC
corresponding to an integrated luminosity of $36.1~\ifb$.
No significant excess of events is
observed above the background expectations. The obtained 95\% CL upper
limits are ${\cal B}\left(H\to\Jpsi\,\gamma\right)<3.5\times10^{-4}$
and ${\cal B}\left(Z\to\Jpsi\,\gamma\right) < 2.3\times10^{-6}$ for
the $\Jpsi\,\gamma$ final state. The corresponding upper limits are ${\cal B}\left(H\to\psi(2S)\,\gamma\right)<2.0\times10^{-3}$
and ${\cal B}\left(Z\to\psi(2S)\,\gamma\right) < 4.5\times10^{-6}$ for
the $\psi(2S)\,\gamma$ final state.
The 95\%~CL upper limits
${\cal B}\left(H\to\Upsilon(nS)\,\gamma\right) < (4.9,5.9,5.7)\times10^{-4}$ and
${\cal B}\left(Z\to\Upsilon(nS)\,\gamma\right) < (2.8,1.7,4.8)\times10^{-6}$
are set for the
$\Upsilon(nS)\,\gamma$ ($n=1,2,3$) final states. These upper limits represent an improvement by a factor of approximately two relative
to the earlier $H(Z)\to\Jpsi\,\gamma$ and $H(Z)\to\Upsilon(nS)\,\gamma$ results from the ATLAS Collaboration using up to
$20.3~\ifb$ of $\sqrt{s}=8~\TeV$ $pp$ collision data with the addition of the first upper limits on the $H/Z\to \psi(2S)\,\gamma$ decays.

\section*{Acknowledgements}


We thank CERN for the very successful operation of the LHC, as well as the
support staff from our institutions without whom ATLAS could not be
operated efficiently.

We acknowledge the support of ANPCyT, Argentina; YerPhI, Armenia; ARC, Australia; BMWFW and FWF, Austria; ANAS, Azerbaijan; SSTC, Belarus; CNPq and FAPESP, Brazil; NSERC, NRC and CFI, Canada; CERN; CONICYT, Chile; CAS, MOST and NSFC, China; COLCIENCIAS, Colombia; MSMT CR, MPO CR and VSC CR, Czech Republic; DNRF and DNSRC, Denmark; IN2P3-CNRS, CEA-DRF/IRFU, France; SRNSFG, Georgia; BMBF, HGF, and MPG, Germany; GSRT, Greece; RGC, Hong Kong SAR, China; ISF and Benoziyo Center, Israel; INFN, Italy; MEXT and JSPS, Japan; CNRST, Morocco; NWO, Netherlands; RCN, Norway; MNiSW and NCN, Poland; FCT, Portugal; MNE/IFA, Romania; MES of Russia and NRC KI, Russian Federation; JINR; MESTD, Serbia; MSSR, Slovakia; ARRS and MIZ\v{S}, Slovenia; DST/NRF, South Africa; MINECO, Spain; SRC and Wallenberg Foundation, Sweden; SERI, SNSF and Cantons of Bern and Geneva, Switzerland; MOST, Taiwan; TAEK, Turkey; STFC, United Kingdom; DOE and NSF, United States of America. In addition, individual groups and members have received support from BCKDF, the Canada Council, CANARIE, CRC, Compute Canada, FQRNT, and the Ontario Innovation Trust, Canada; EPLANET, ERC, ERDF, FP7, Horizon 2020 and Marie Sk{\l}odowska-Curie Actions, European Union; Investissements d'Avenir Labex and Idex, ANR, R{\'e}gion Auvergne and Fondation Partager le Savoir, France; DFG and AvH Foundation, Germany; Herakleitos, Thales and Aristeia programmes co-financed by EU-ESF and the Greek NSRF; BSF, GIF and Minerva, Israel; BRF, Norway; CERCA Programme Generalitat de Catalunya, Generalitat Valenciana, Spain; the Royal Society and Leverhulme Trust, United Kingdom.

The crucial computing support from all WLCG partners is acknowledged gratefully, in particular from CERN, the ATLAS Tier-1 facilities at TRIUMF (Canada), NDGF (Denmark, Norway, Sweden), CC-IN2P3 (France), KIT/GridKA (Germany), INFN-CNAF (Italy), NL-T1 (Netherlands), PIC (Spain), ASGC (Taiwan), RAL (UK) and BNL (USA), the Tier-2 facilities worldwide and large non-WLCG resource providers. Major contributors of computing resources are listed in Ref.~\cite{ATL-GEN-PUB-2016-002}.

\printbibliography
\clearpage
 
\begin{flushleft}
{\Large The ATLAS Collaboration}

\bigskip

M.~Aaboud$^\textrm{\scriptsize 34d}$,    
G.~Aad$^\textrm{\scriptsize 99}$,    
B.~Abbott$^\textrm{\scriptsize 124}$,    
O.~Abdinov$^\textrm{\scriptsize 13,*}$,    
B.~Abeloos$^\textrm{\scriptsize 128}$,    
D.K.~Abhayasinghe$^\textrm{\scriptsize 91}$,    
S.H.~Abidi$^\textrm{\scriptsize 164}$,    
O.S.~AbouZeid$^\textrm{\scriptsize 39}$,    
N.L.~Abraham$^\textrm{\scriptsize 153}$,    
H.~Abramowicz$^\textrm{\scriptsize 158}$
% ...
\bigskip
\\

% ...
$^{13}$Institute of Physics, Azerbaijan Academy of Sciences, Baku; Azerbaijan.\\
% ...
$^{34}$$^{(a)}$Facult\'e des Sciences Ain Chock, R\'eseau Universitaire de Physique des Hautes Energies - Universit\'e Hassan II, Casablanca;$^{(b)}$Centre National de l'Energie des Sciences Techniques Nucleaires (CNESTEN), Rabat;$^{(c)}$Facult\'e des Sciences Semlalia, Universit\'e Cadi Ayyad, LPHEA-Marrakech;$^{(d)}$Facult\'e des Sciences, Universit\'e Mohamed Premier and LPTPM, Oujda;$^{(e)}$Facult\'e des sciences, Universit\'e Mohammed V, Rabat; Morocco.\\
% ...
$^{39}$Niels Bohr Institute, University of Copenhagen, Copenhagen; Denmark.\\
% ...
$^{91}$Department of Physics, Royal Holloway University of London, Egham; United Kingdom.\\
% ...
$^{99}$CPPM, Aix-Marseille Universit\'e, CNRS/IN2P3, Marseille; France.\\
% ...
$^{124}$Homer L. Dodge Department of Physics and Astronomy, University of Oklahoma, Norman OK; United States of America.\\
% ...
$^{128}$LAL, Universit\'e Paris-Sud, CNRS/IN2P3, Universit\'e Paris-Saclay, Orsay; France.\\
% ...
$^{153}$Department of Physics and Astronomy, University of Sussex, Brighton; United Kingdom.\\
% ...
$^{158}$Raymond and Beverly Sackler School of Physics and Astronomy, Tel Aviv University, Tel Aviv; Israel.\\
% ...
$^{164}$Department of Physics, University of Toronto, Toronto ON; Canada.\\
% ...
$^{a}$ Also at Borough of Manhattan Community College, City University of New York, NY; United States of America.\\
$^{b}$ Also at California State University, East Bay; United States of America.\\
$^{c}$ Also at Centre for High Performance Computing, CSIR Campus, Rosebank, Cape Town; South Africa.\\
$^{d}$ Also at CERN, Geneva; Switzerland.\\
$^{e}$ Also at CPPM, Aix-Marseille Universit\'e, CNRS/IN2P3, Marseille; France.\\
% ...
$^{*}$ Deceased

\end{flushleft}


\end{document}